%% file: main.tex
\pdfoutput=1
\documentclass[sn-mathphys-num]{sn-jnl}% Math and Physical Sciences Numbered Reference Style 
\usepackage{graphicx}%
\usepackage{multirow}%
\usepackage{amsmath,amssymb,amsfonts}%
\usepackage{amsthm}%
\usepackage{mathrsfs}%
\usepackage[title]{appendix}%
\usepackage{xcolor}%
\usepackage{textcomp}%
\usepackage{manyfoot}%
\usepackage{booktabs}%
\usepackage{algorithm}%
\usepackage{algorithmicx}%
\usepackage{algpseudocode}%
\usepackage{listings}%

\definecolor{codebg}{RGB}{240,240,240} % Light gray background
\definecolor{codeframe}{RGB}{180,180,180} % Frame color
\definecolor{codefont}{RGB}{0,0,0} % Black text for better readability

\lstdefinestyle{mystyle}{
    backgroundcolor=\color{codebg},   
    keywordstyle=\color{magenta},
    numberstyle=\tiny\color{gray},
    basicstyle=\ttfamily\footnotesize,
    breakatwhitespace=false,         
    breaklines=true,                 
    captionpos=b,                    
    keepspaces=true,                 
    numbers=left,                    
    numbersep=5pt,                  
    showspaces=false,                
    showstringspaces=false,
    showtabs=false,                  
    tabsize=2
}

\newcommand{\noncopynumber}[1]{
    \BeginAccSupp{method=escape,ActualText={}}
    #1
    \EndAccSupp{}
}

\usepackage{accsupp}    
\usepackage{fontawesome5}
\usepackage{tikz}
\usetikzlibrary{shapes.geometric, shapes.misc, arrows.meta, positioning}
\usepackage{subcaption} % For subfigures
\usepackage{tikz}
\usetikzlibrary{shapes.geometric, arrows, positioning, fit, calc}

\tikzstyle{startstop} = [rectangle, rounded corners, minimum width=3cm, minimum height=1cm,text centered, draw=black, fill=red!30]
\tikzstyle{process} = [rectangle, minimum width=3cm, minimum height=1cm, text centered, draw=black, fill=blue!30]
\tikzstyle{decision} = [diamond, minimum width=3cm, minimum height=1cm, text centered, draw=black, fill=green!30]
\tikzstyle{arrow} = [thick,->,>=stealth]
\tikzstyle{data} = [trapezium, trapezium left angle=60, trapezium right angle=120, minimum width=3cm, minimum height=1cm, text centered, draw=black, fill=yellow!30]
\tikzstyle{database} = [cylinder, draw=black, shape border rotate=90, text centered, minimum height=1cm, minimum width=1cm, fill=orange!30]

\theoremstyle{thmstyleone}%
\theoremstyle{thmstyletwo}%

\theoremstyle{thmstylethree}%

\begin{document}

\title[Advanced Applications of Generative AI in Actuarial Science: Case Studies Beyond ChatGPT]{Advanced Applications of Generative AI in Actuarial Science: Case Studies Beyond ChatGPT}


\author*[1]{\fnm{Simon} \sur{Hatzesberger}}\email{simon.hatzesberger@gmail.com}
\equalcont{These authors contributed equally to this work.}

\author[2]{\fnm{Iris} \sur{Nonneman}}
\equalcont{These authors contributed equally to this work.}

\affil[1]{\orgdiv{Actuarial \& Insurance Services}, \orgname{Deloitte}, \orgaddress{\country{Germany}}}

%\affil{\orgdiv{TODO}, \orgname{Achmea}, \orgaddress{\country{Netherlands}}}


\abstract{
This article explores the potential of generative AI (GenAI) to support actuarial practice through four implemented case studies.
It situates these case studies within the broader evolution of artificial intelligence in actuarial science, from early neural networks and machine learning to modern transformer-based GenAI systems.
The first case study illustrates how large language models (LLMs) can improve claim cost prediction by extracting informative features from unstructured text for use in the underlying supervised learning task.
The second case study demonstrates the automation of market comparisons using Retrieval-Augmented Generation to identify, extract, and structure relevant information from insurers' annual reports.
The third case study highlights the capabilities of fine-tuned vision-enabled LLMs in classifying car damage types and extracting contextual information from images.
The fourth case study presents a multi-agent system that autonomously migrates actuarial legacy code from R to Python and validates the translation against the original code's outputs.
In addition to these case studies, we outline further GenAI applications in the insurance industry. 
Finally, we discuss the regulatory, security, dual-use and fraud, reproducibility, privacy, governance, and organisational challenges associated with deploying GenAI in regulated insurance environments.
}

\keywords{Generative AI, Case Studies, Large Language Models, Retrieval-Augmented Generation, Fine-Tuning, Multi-Agent Systems}



\maketitle

%%%%%%%%%%%%%%%%%%%%%%%%%%%%%%%%%%%%%%%%%%%%%%%%%%%
% Section 1: Introduction
%%%%%%%%%%%%%%%%%%%%%%%%%%%%%%%%%%%%%%%%%%%%%%%%%%%
\input{introduction}

%%%%%%%%%%%%%%%%%%%%%%%%%%%%%%%%%%%%%%%%%%%%%%%%%%%
% Section 2: Background and Context
%%%%%%%%%%%%%%%%%%%%%%%%%%%%%%%%%%%%%%%%%%%%%%%%%%%
\input{background_and_context}

%%%%%%%%%%%%%%%%%%%%%%%%%%%%%%%%%%%%%%%%%%%%%%%%%%%
% Section 3: Case Study 1
%%%%%%%%%%%%%%%%%%%%%%%%%%%%%%%%%%%%%%%%%%%%%%%%%%%
\input{case_study_crash_reports}

%%%%%%%%%%%%%%%%%%%%%%%%%%%%%%%%%%%%%%%%%%%%%%%%%%%
% Section 4: Case Study 2
%%%%%%%%%%%%%%%%%%%%%%%%%%%%%%%%%%%%%%%%%%%%%%%%%%%
\input{case_study_market_comparison}

%%%%%%%%%%%%%%%%%%%%%%%%%%%%%%%%%%%%%%%%%%%%%%%%%%%
% Section 5: Case Study 3
%%%%%%%%%%%%%%%%%%%%%%%%%%%%%%%%%%%%%%%%%%%%%%%%%%%
\input{case_study_vision_fine_tuning}

%%%%%%%%%%%%%%%%%%%%%%%%%%%%%%%%%%%%%%%%%%%%%%%%%%%
% Section 6: Case Study 4
%%%%%%%%%%%%%%%%%%%%%%%%%%%%%%%%%%%%%%%%%%%%%%%%%%%
\input{case_study_multi_agent_system}

%%%%%%%%%%%%%%%%%%%%%%%%%%%%%%%%%%%%%%%%%%%%%%%%%%%
% Section 7: Further Applications of Generative AI
%%%%%%%%%%%%%%%%%%%%%%%%%%%%%%%%%%%%%%%%%%%%%%%%%%%
\input{further_applications}

%%%%%%%%%%%%%%%%%%%%%%%%%%%%%%%%%%%%%%%%%%%%%%%%%%%
% Section 8: Risks and Governance of GenAI in Insurance
%%%%%%%%%%%%%%%%%%%%%%%%%%%%%%%%%%%%%%%%%%%%%%%%%%%
\input{challenges_and_considerations}

%%%%%%%%%%%%%%%%%%%%%%%%%%%%%%%%%%%%%%%%%%%%%%%%%%%
% Section 9: Conclusion
%%%%%%%%%%%%%%%%%%%%%%%%%%%%%%%%%%%%%%%%%%%%%%%%%%%
\input{conclusion}

%%%%%%%%%%%%%%%%%%%%%%%%%%%%%%%%%%%%%%%%%%%%%%%%%%%
% Backmatter
%%%%%%%%%%%%%%%%%%%%%%%%%%%%%%%%%%%%%%%%%%%%%%%%%%%

\backmatter

\bmhead{Supplementary information}

All materials related to the case studies presented in this paper -- including datasets, Jupyter notebooks containing the full source code, and explicit listings of package dependencies with pinned version specifications for compatibility -- are available in dedicated subfolders of the GitHub repository at \url{https://github.com/IAA-AITF/Actuarial-AI-Case-Studies/}. To ensure exact reproducibility, all results reported in this article correspond to the tagged release \texttt{eaj-v1.0} (commit \texttt{af16c5b}) of this repository, and each case-study folder provides a \texttt{requirements.txt} file that pins the exact package versions used. These resources facilitate passive review of the code and its intermediate results, as well as active experimentation, enabling readers to run, modify, and extend the implementations within their own computational environments.

\bmhead{Acknowledgements}

We sincerely thank the Young Actuaries Initiative and the AI Task Force of the International Actuarial Association for bringing us together, broadening our perspectives, enhancing our skills, and fostering our growth and collaboration in this field.

\bmhead{Funding}

The authors did not receive support from any organisation for the submitted work.

\bmhead{Conflict of interest}

The authors have no competing interests to declare that are relevant to the content of this article.

%\section*{Declarations}
%
%Some journals require declarations to be submitted in a standardised format. Please check the Instructions for Authors of the journal to which you are submitting to see if you need to complete this section. If yes, your manuscript must contain the following sections under the heading `Declarations':
%
%\begin{itemize}
%\item Conflict of interest/Competing interests (check journal-specific guidelines for which heading to use)
%\item Ethics approval and consent to participate
%\item Consent for publication
%\item Data availability 
%\end{itemize}
%
%\noindent
%If any of the sections are not relevant to your manuscript, please include the heading and write `Not applicable' for that section. 
%
%%%===================================================%%
%%% For presentation purpose, we have included        %%
%%% \bigskip command. Please ignore this.             %%
%%%===================================================%%
%\bigskip
%\bigskip\noindent
%Springer journals and proceedings: \url{https://www.springer.com/gp/editorial-policies}
%
%\bigskip\noindent
%Nature Portfolio journals: \url{https://www.nature.com/nature-research/editorial-policies}
%
%\bigskip\noindent
%\textit{Scientific Reports}: \url{https://www.nature.com/srep/journal-policies/editorial-policies}
%
%\bigskip\noindent
%BMC journals: \url{https://www.biomedcentral.com/getpublished/editorial-policies}
%\end{flushleft}
%
\bibliography{sn-bibliography}% common bib file

\end{document}

%% file: introduction.tex
\section{Introduction}
\label{sec:introduction}

The rapid development of artificial intelligence (AI) is changing many fields, and actuarial science and insurance are no exception. Traditionally, actuarial work relied on statistical modelling and historical data analysis to assess risks and make informed decisions. The landscape has shifted as actuaries have adopted machine learning and deep learning methods into their analytical frameworks \cite{richman2021ai, richman2021_2, Wuthrich2025AITools}.
Building on these advances, the emergence of generative AI (GenAI) has introduced further approaches that can support actuarial work. GenAI offers opportunities to improve predictive accuracy, streamline operational processes, and unlock valuable insights from unstructured data sources throughout the entire insurance value chain \cite{carlin2024primer, balona2023actuarygpt, eling2022impact}.
While GenAI holds considerable potential in actuarial and insurance contexts, the very capabilities that make it valuable also introduce new risks -- a so-called dual-use nature, whereby the same algorithms can be exploited for both beneficial and harmful purposes \cite{koplin2023}. Against this backdrop, this article explores both the current and potential impact of GenAI on actuarial practice, encouraging its thoughtful integration into operational workflows. Through four case studies, we show how GenAI methods can be applied to actuarial work.

The article is structured as follows. Section~\ref{sec:background_and_context} provides background and context, tracing the evolution of AI -- from early neural networks (NNs) to modern GenAI systems -- and discussing its implications for actuarial science. Sections~\ref{sec:case_study_crash_reports} to \ref{sec:multi_agent_system} present four fully implemented case studies, each illustrating practical applications of GenAI to real-world actuarial problems: enhancing claim cost prediction by using large language models (LLMs) to derive predictive features from unstructured claim descriptions, automating structured market comparisons based on insurers' annual reports, classifying and localising car damages from images using vision-enabled models, and using a multi-agent system to migrate actuarial legacy code from R to Python while validating the translated outputs against the original code. Section~\ref{sec:further_applications} explores additional applications of GenAI beyond these case studies. Finally, Section~\ref{sec:challenges_and_considerations} discusses the key risks and governance considerations associated with deploying GenAI in insurance, including regulatory, security, dual-use and fraud, reproducibility, privacy, governance, and organisational aspects.

The case studies serve a dual purpose. First, they provide an educational foundation, demonstrating what is already achievable today with GenAI in actuarial contexts. Second, they are intended to inspire actuaries to integrate GenAI into their workflows by showcasing concrete use cases with immediate practical relevance. All four case studies have been implemented in Jupyter notebooks, which are made available as supplementary material on the GitHub account of the International Actuarial Association\footnote{\url{https://github.com/IAA-AITF/Actuarial-AI-Case-Studies}}. The notebooks are designed to run end-to-end and can be copied or adapted as templates for readers' own actuarial GenAI projects. With this article, we aim to equip actuaries with the knowledge and practical resources to integrate GenAI responsibly into their own work.

%% file: background_and_context.tex
\section{Background and Context}
\label{sec:background_and_context}

\subsection{Historical Perspective on the Use of AI in Actuarial Science}

Artificial intelligence has become an increasingly influential force in the world economy. With the rise of transformer-based models \cite{vaswani2017attention} -- particularly large language models -- the scope of AI applications has expanded beyond structured data formats like tabular data to include unstructured data such as text and images. Widely adopted GenAI tools such as ChatGPT and Gemini have introduced these models to a broad audience in the form of practical chatbots.

Although the adoption of LLMs has been relatively recent, AI has its origins in the second half of the 20th century. With the expansion of computing capacity, researchers developed new algorithms capable of performing large-scale, complex mathematical calculations.
In the 1980s, the introduction of backpropagation \cite{rumelhart1986learning} advanced NN research. By the end of the century, sophisticated NNs and decision trees had been developed alongside a shift towards probabilistic methods and data-driven models. This new focus led to the development of support vector machines (SVMs) and other data-driven methods, which found applications across diverse domains including natural language processing (NLP) and computer vision.
The 2010s saw rapid developments in deep learning: the introduction of AlexNet (2012) for image recognition~\cite{krizhevsky2012imagenet}, word embedding techniques such as GloVe (2014)~\cite{pennington2014glove}, and transformer models (2017)~\cite{vaswani2017attention} unlocked potential for working with large, unstructured, and non-tabular datasets. Combined with the increase in computational resources, these models drove a widespread interest in the adoption of AI, marking a notable shift from previous decades when limited data availability and computational constraints hindered most sectors.
The actuarial profession faced similar limitations throughout the late 20th century, leading actuaries to rely primarily on the usage of practical algorithms like Generalised Linear Models (GLMs) \cite{nelder1972generalized} and analytical methods like the Bornhuetter-Ferguson technique for Incurred But Not Reported (IBNR) reserving \cite{bornhuetter1972ibnr}. As computing power and data availability grew, AI algorithms were gradually adopted in actuarial science. One of the first studies analysed several machine learning methods in comparison to GLMs \cite{dugas2003statistical}. Subsequent work compared the predictive power of GLMs against NNs~\cite{wuthrich2019glms}, while further studies explored the potential of tree-based methods \cite{denuit2020effective, henckaerts2021boosting, wuthrich2023data}. Actuaries have also begun exploring transformer models, such as the Credibility Transformer \cite{richman2024credibility}, in actuarial applications.

This growing integration of AI in actuarial science aligns with the perspective of Charles Cowling, past president of the International Actuarial Association, who has repeatedly emphasised that ``AI will not replace actuaries, but actuaries with AI will replace actuaries without AI'' \cite{cowling2024ai}.

\subsection{Current Trends and Advancements in Generative AI Technologies}

With the introduction of transformer models, GenAI has emerged as a rapidly developing area within the field of artificial intelligence. While traditional AI models focus on predicting outcomes based on input data, GenAI uses the provided data to create new information. For instance, image models now move beyond classifying images into categories to generating entirely new images for these categories. 
Although GenAI can be applied across various data modalities, its most notable advancements have occurred in the domain of textual data. GenAI has advanced NLP, elevating it to a central field of research and innovation. State-of-the-art LLMs, such as OpenAI's GPT series and Anthropic's Claude models, showcase the advanced capabilities of this technology. Their functionality ranges from basic question answering to powering advanced, multi-turn conversational agents that facilitate complex interactions across a wide range of applications.
Following the early success of transformer models in NLP, GenAI research has expanded into domains such as video generation~\cite{brooks2024sora}, audio synthesis~\cite{borsos2023audiolm}, and applications in medicine~\cite{lobentanzer2025platform}. The launch of reasoning-oriented models such as OpenAI's o1 series in 2024 marks a further advancement, enabling LLMs to tackle complex reasoning tasks across disciplines such as mathematics, biology, and chemistry with enhanced precision. In parallel, AI-assisted code generation has advanced, with coding agents now capable of autonomously resolving non-trivial software engineering tasks~\cite{huynh2025codegen, jimenez2024swebench}. The emergence of agentic AI has introduced a new approach, enabling autonomous collaboration among specialised models to address complex, multi-step workflows. Early research has also explored computer-use systems that interpret graphical user interfaces via screenshots and operate them through simulated mouse and keyboard actions, pointing towards broader task automation beyond text.

\subsection{The Role of Generative AI in Transforming Actuarial Work}

GenAI is a promising technology with the potential to transform business processes, including those within actuarial work~\cite{carlin2024primer, balona2023actuarygpt}. 
The actuarial ecosystem, characterised by vast data volumes and complex calculations, is well-positioned to benefit from GenAI. It can assist actuaries by streamlining workflows related to risk assessment, extracting insights from unstructured data, and supporting granular risk analyses. For example, in insurance underwriting, multimodal LLMs can rapidly synthesise textual data alongside tabular policyholder data and medical reports.
On the other hand, LLMs can also be used to create more accurate predictions. For example, the Credibility Transformer~\cite{richman2024credibility}, which extends the transformer architecture with a credibility mechanism, can achieve higher predictive accuracy than traditional deep learning models. 
Moreover, a framework has been proposed for modelling claim frequency and loss severity that uses Bidirectional Encoder Representations from Transformers (BERT) to extract descriptive textual information from claim records, with predictions generated via NNs~\cite{xu2023bert}.
Beyond improved predictions, AI also has the potential to automate operational workflows~\cite{richman2024ai, yetistiren2023evaluating}.
This opportunity is further acknowledged by \cite{eling2022impact}, whose research highlights GenAI's potential for process automation and customer personalisation, while also cautioning that its introduction changes the risk landscape for the insurance industry. Risks arise, ranging from deepfakes \cite{mitra2024}, harmful content generation, and copyright concerns \cite{franceschelli2022copyright} to broader risks for businesses and insurers \cite{Geneva2025}. Some of these risks are novel, while others amplify existing ones. Within actuarial science, GenAI can be exploited to generate convincing narratives, fabricated documents, and synthetic images, enabling novel forms of insurance fraud \cite{swissre2025sonar}. On the business side, GenAI can cause tangible harm when applied incorrectly, particularly in safety-critical contexts \cite{oviedotrespalacios2023risks}. In actuarial practice, where the quality and correctness of outputs carry significant professional, reputational, and regulatory weight, LLM-generated content must always be subject to expert review and validation, supported by risk assessments, ethical guidelines, and careful governance. Responsible adoption of GenAI therefore requires developing awareness of its adversarial potential and embedding robust validation frameworks into everyday workflows. 
The following sections present and analyse the potential of GenAI through four actuarial case studies, illustrating practical applications and demonstrating how this technology can be applied in real-world actuarial work.

%% file: case_study_crash_reports.tex
\section{Case Study 1: Improving Claim Cost Prediction with LLM-Extracted Features from Unstructured Data}
\label{sec:case_study_crash_reports}

\subsection{Introduction}

The insurance industry has long faced challenges in fully leveraging unstructured text data that contains valuable insights for claims assessment and risk management. Traditional analytical methods for claim cost prediction primarily focus on structured tabular data, thereby overlooking critical details embedded in texts such as claim descriptions, incident reports, and customer communications.

In this case study, we show how LLMs can be employed to transform unstructured textual data into structured, actionable information. The goal of this case study is to generate new features from claim descriptions that can be used in a machine learning model to predict ultimate incurred claim costs. By enhancing our model with additional features derived from claim descriptions, we aim to improve its predictive accuracy, while also gaining deeper insight into the factors that influence claim size.

To achieve this, we examine a workers' compensation claims dataset which consists of tabular data and additional textual claim descriptions. We demonstrate how LLMs can extract key information such as injured body parts and accident causes from these descriptions. The extracted information is then used to construct new features for a gradient boosting model tasked with predicting claim costs. Our work extends a line of actuarial research that has progressively incorporated NLP into structured claim cost models, from word-embedding features~\cite{lee2020wordembeddings} to BERT-based severity and frequency models~\cite{xu2022bertwarranty, xu2023bert}.

The complete case study has been implemented in a Jupyter notebook and is accessible on GitHub\footnote{\url{https://github.com/IAA-AITF/Actuarial-AI-Case-Studies/tree/eaj-v1.0/case-studies/2025/claim_cost_prediction_with_LLM-extracted_features}}, where additional technical details can be found.

\subsection{Approach and Techniques}

\subsubsection*{Data and Baseline Model}

For this case study, we use a dataset of 3,000 workers' compensation claims from a Kaggle competition on actuarial loss estimation\footnote{\url{https://www.kaggle.com/competitions/actuarial-loss-estimation}}. This dataset is fully synthetic -- constructed as a stratified sample from a larger pool of 90,000 realistically simulated records -- and each record represents a unique claim. It includes both structured features (e.g., age, gender, marital status, wages) and unstructured text descriptions of the claim reports.

As stated in the previous subsection, our aim is to enhance a prediction model with features derived by an LLM and assess the performance of this enhanced model compared to a baseline model. To this end, we first create a baseline model, which uses existing tabular data as features to predict the ultimate incurred claim costs. For this task, we train a gradient boosting regression model.

\subsubsection*{LLM-Based Feature Extraction}

Next, we employ an LLM to analyse the claim descriptions and extract structured information, in particular the primary body part injured, the cause of injury, and the number of body parts involved. Specifically, we use OpenAI's GPT-4o mini (version identifier: \texttt{gpt-4o-mini-2024-07-18}) via the Structured Outputs API, which ensures that the model reliably returns the three target fields in a structured format. The following system prompt is applied iteratively to each claim description to extract new features:

\begin{lstlisting}[language=]
prompt = """
    Your task is to extract structured information about
    injuries and cause of injury from the given text.
    Follow this schema strictly:
    - number_of_body_parts_injured: The total count of injured body parts.
    - main_body_part_injured: The primary body part affected, described
      concisely (e.g., 'HEAD', 'THUMB').
    - cause_of_injury: Specify by verb.
        - If verb given: return only the primary action verb that directly
          caused the injury (e.g. 'fall' not 'fell from box')
        - If cause is not mentioned, infer from context if possible,
          otherwise return 'unspecified'.
    Ensure accuracy and consistency in the extracted details.
    Do not add interpretations beyond the provided text.
"""
\end{lstlisting}

Every output contains structured information corresponding to the three fields above, which are formatted into the additional features \texttt{number\_of\_body\_parts\_injured}, \texttt{main\_body\_part\_injured}, and \texttt{cause\_of\_injury}. Raw LLM outputs (so-called completions) are cached to ensure downstream reproducibility, since LLMs can exhibit run-to-run variance even when the temperature parameter is set to~$0.0$ (see Section~\ref{sec:nondeterminism}).

While the Structured Outputs API guarantees that every response adheres to the prescribed schema, the extracted string values themselves remain free-form and yield a large number of distinct categories for injured body parts and cause of injury. To make these features tractable for downstream modelling, we group the raw LLM outputs into broader categories using a deterministic codebook: a fixed set of regular-expression rules that maps each output string to one of 8 anatomical regions (e.g., \texttt{TORSO}, \texttt{HAND\_FINGERS}) or one of 13 cause categories (e.g., \texttt{LIFTING\_CARRYING}, \texttt{IMPACT}, \texttt{FALL\_SLIP\_TRIP}), yielding the new categorical features \texttt{body\_part\_category} and \texttt{cause\_of\_injury\_category}. The codebook was developed through domain-expert review, uses no statistical parameters derived from the data, and is documented in full in the accompanying notebook. More elaborate dimensionality-reduction alternatives, such as entity embeddings, offer a natural extension that we leave to the reader.

To evaluate extraction quality, the two authors jointly annotated a random sample of 100 claims by consensus, assigning gold-standard labels for body-part category and cause-of-injury category following the same codebook used for the LLM outputs. Because this consensus procedure yields a single agreed label per claim, we do not report an inter-annotator agreement statistic; the modest sample size and the single-pass consensus labelling are acknowledged limitations of this extraction-quality check. The resulting field-level accuracies are reported in the Results subsection below.

After creating these additional features, we construct an enhanced gradient boosting regression model. Compared to the baseline model, this model not only incorporates the structured features used previously but also includes the LLM-extracted features. Before fitting both models, we apply a log transformation to the target variable \texttt{UltimateIncurredClaimCost} to address distribution skewness.

\subsubsection*{Evaluation Techniques}
We employ an evaluation framework designed to produce stable performance estimates and to quantify their uncertainty. The data are split into a training set (80\%) and a held-out test set (20\%) using stratified sampling on deciles of the log-transformed target variable. Stratification ensures that each part of the claim cost distribution is proportionally represented in both splits, which is particularly important for the stratified performance analysis reported below. The test set is only used for the final performance evaluation. Stratified 4-fold cross-validation (CV) on the training set is used both for hyperparameter selection and to quantify the stability of performance estimates; the two procedures use different random seeds so that the evaluation folds are independent of the folds that informed hyperparameter selection.

The LLM-derived categorical features \texttt{body\_part\_category} and \texttt{cause\_of\_injury\_category} are one-hot encoded prior to splitting the data. Because the category set is fully determined by the deterministic codebook rather than discovered from the training data, the column structure is known in advance and no unseen category can arise in the test set.

Hyperparameter selection over the number of trees (\texttt{n\_estimators}) and maximum tree depth (\texttt{max\_depth}) is performed via Stratified 4-fold CV on the training set, evaluating a grid of nine combinations ($\texttt{n\_estimators} \in \{50, 100, 200\}$, $\texttt{max\_depth} \in \{3, 5, 7\}$) and selecting the combination that minimises mean CV Root Mean Squared Error (RMSE). This procedure is applied independently to the baseline and the enhanced model, and to each variant in the ablation study below.

We evaluate model performance using the RMSE, the Mean Absolute Error (MAE), and the $R^2$ score. In the notebook, we additionally report quantile (pinball) loss at the 50th, 75th, and 90th percentiles, which captures upper-tail accuracy. Finally, we examine the gradient boosting model's built-in feature importance (mean impurity decrease) to assess the most influential predictors.

\subsection{Results}

\subsubsection*{LLM Extraction Quality}
The raw LLM outputs are reduced from 224 unique values for \texttt{main\_body\_part\_injured} and 175 for \texttt{cause\_of\_injury} to 8 and 13 categories, respectively, through the deterministic codebook described above. On the 100-claim gold-standard sample, the LLM achieves 91\% category-level accuracy for body-part classification and 83\% for cause-of-injury classification, confirming that GPT-4o mini extracts meaningful and consistent information from the claim descriptions.

\subsubsection*{Overall Performance}
Table~\ref{tab:performance_comparison} presents the main performance comparison. Both CV estimates from the Stratified 4-fold CV procedure and the held-out test evaluation are reported. For each metric, the mean and standard deviation (SD) across the 4 CV folds quantify the stability of the improvement.

\begin{table}[h!]
\centering
\caption{Performance comparison between baseline model and enhanced model on Stratified 4-fold CV (training set) and the held-out test set. Metrics marked with ($\downarrow$) are better when lower; ($\uparrow$) are better when higher.}
\label{tab:performance_comparison}
\begin{tabular}{lccc}
\toprule
\textbf{Metric} & \textbf{Baseline} & \textbf{Enhanced} & \textbf{Improvement} \\
\midrule
\multicolumn{4}{l}{\textit{CV (mean $\pm$ SD, 4 folds)}} \\[2pt]
RMSE $(\downarrow)$ & $1.321 \pm 0.020$ & $1.113 \pm 0.025$ & 15.8\% \\
MAE  $(\downarrow)$ & $1.085 \pm 0.024$ & $0.861 \pm 0.023$ & 20.6\% \\
$R^2$ $(\uparrow)$  & $0.245 \pm 0.017$ & $0.465 \pm 0.024$ & 89.4\% \\
\midrule
\multicolumn{4}{l}{\textit{Held-out test set}} \\[2pt]
RMSE $(\downarrow)$ & 1.345 & 1.101 & 18.2\% \\
MAE  $(\downarrow)$ & 1.078 & 0.834 & 22.6\% \\
$R^2$ $(\uparrow)$  & 0.224 & 0.481 & 114.5\% \\
\bottomrule
\end{tabular}
\end{table}

On the held-out test set, the enhanced model achieves an 18.2\% reduction in RMSE (from 1.345 to 1.101), a 22.6\% reduction in MAE, and an increase in $R^2$ from 0.224 to 0.481. The large relative improvement in $R^2$ reflects the low absolute baseline value rather than a disproportionate absolute gain. The CV results confirm that these improvements are stable: the mean RMSE reduction across 4 folds is 0.209 (15.8\%). This stability extends beyond the single fold split: re-running the evaluation with ten different random 4-fold partitions, and likewise under 5-fold and 10-fold cross-validation (ten partitions each), leaves the RMSE, MAE, and $R^2$ gains essentially unchanged (standard deviations below $0.01$ across the thirty repetitions), and the enhanced model outperforms the baseline in every one of them.

\subsubsection*{Statistical Significance}
To verify that the observed performance gain is not attributable to chance, we conduct a formal significance test. Because CV fold scores share training data, the standard paired $t$-test understates variance and can be anti-conservative. We therefore apply the corrected paired $t$-test of \cite{nadeau2003inference}, which inflates the standard error by the factor $\sqrt{1/K + n_\mathrm{test}/n_\mathrm{train}}$ to account for the non-independence of folds. For RMSE the corrected test yields $t = 14.51$, $p = 7.1 \times 10^{-4}$; for MAE it yields $t = 18.37$, $p = 3.5 \times 10^{-4}$ (both with $K{-}1 = 3$ degrees of freedom). Both results are significant at $\alpha = 0.05$. We conclude that the observed improvement on this dataset and evaluation protocol is statistically significant, though independent replication on a larger or different dataset would be needed to establish broader generalisability.

\subsubsection*{Ablation Study}
To attribute the performance lift to specific LLM-derived feature groups, Table~\ref{tab:ablation} presents results for five feature variants evaluated under the same data splits; each variant's hyperparameters were selected independently via the same tuning procedure. The results show that the body-part categories contribute the largest individual gain (RMSE $1.135 \pm 0.023$), followed by the cause-of-injury categories (RMSE $1.184 \pm 0.036$). The injury-count feature alone provides negligible lift (RMSE $1.321 \pm 0.023$), essentially identical to the baseline. Combining all three groups yields the best performance (RMSE $1.113 \pm 0.025$), confirming that the body-part and cause-of-injury features are complementary.

\begin{table}[h!]
\centering
\caption{Ablation study: performance of individual LLM-derived feature groups and their full combination, each compared to the tabular-only baseline. Results are mean $\pm$ SD over Stratified 4-fold CV on the training set. Best parameters refer to the selected \texttt{n\_estimators} and \texttt{max\_depth}.}
\label{tab:ablation}
\begin{tabular}{lcccc}
\toprule
\textbf{Variant} & \textbf{Best params} & \textbf{RMSE} & \textbf{MAE} & \boldmath{$R^2$} \\
& & \footnotesize{mean $\pm$ SD} & \footnotesize{mean $\pm$ SD} & \footnotesize{mean $\pm$ SD} \\
\midrule
Baseline (tabular only) & (50, 3) & $1.321 \pm 0.020$ & $1.085 \pm 0.024$ & $0.245 \pm 0.017$ \\
+ Body-part categories & (200, 3) & $1.135 \pm 0.023$ & $0.883 \pm 0.023$ & $0.443 \pm 0.021$ \\
+ Cause-of-injury categories & (100, 3) & $1.184 \pm 0.036$ & $0.934 \pm 0.027$ & $0.393 \pm 0.037$ \\
+ Injury-count feature & (50, 3) & $1.321 \pm 0.023$ & $1.085 \pm 0.024$ & $0.246 \pm 0.020$ \\
+ All LLM-derived features & (200, 3) & $1.113 \pm 0.025$ & $0.861 \pm 0.023$ & $0.465 \pm 0.024$ \\
\bottomrule
\end{tabular}
\end{table}

\subsubsection*{Stratified Evaluation by Claim Cost}
Aggregate metrics can mask heterogeneous performance across different claim sizes. Table~\ref{tab:stratified} reports RMSE and MAE on the held-out test set stratified by original (non-log) claim cost band. The enhanced model improves over the baseline in three of the four bands, with the improvement most pronounced in the extreme bands: for claims below \$1,000 the RMSE falls from 1.607 to 1.059, and for claims above \$20,000 it falls from 2.371 to 2.085. In the \$1k--\$5k band the two models perform comparably (RMSE 0.745 vs.\ 0.785).

\begin{table}[h!]
\centering
\caption{Stratified evaluation on the held-out test set by original (non-log) claim cost band. All metrics are computed on the log-transformed scale used for modelling.}
\label{tab:stratified}
\begin{tabular}{lrcccc}
\toprule
& & \multicolumn{2}{c}{\textbf{RMSE}} & \multicolumn{2}{c}{\textbf{MAE}} \\
\cmidrule(lr){3-4} \cmidrule(lr){5-6}
\textbf{Band} & \textbf{Number of Rows} & \textbf{Baseline} & \textbf{Enhanced} & \textbf{Baseline} & \textbf{Enhanced} \\
\midrule
$<$\,\$1k       & 155 & 1.607 & 1.059 & 1.492 & 0.888 \\
\$1k--\$5k      & 194 & 0.745 & 0.785 & 0.594 & 0.628 \\
\$5k--\$20k     & 179 & 0.983 & 0.821 & 0.809 & 0.612 \\
$>$\,\$20k      &  72 & 2.371 & 2.085 & 2.160 & 1.824 \\
\bottomrule
\end{tabular}
\end{table}

\subsubsection*{Feature Importance}
To understand which predictors drive the model's predictions and to support domain validation, we examine the gradient boosting model's built-in feature importance based on mean impurity decrease. Among the top-ranked predictors are both structured features and LLM-derived categories. \texttt{WeeklyWages} and \texttt{Age} consistently rank highest among the structured features. Among the LLM-derived features, \texttt{body\_part\_category} (in particular \texttt{TORSO} and \texttt{HAND\_FINGERS}) and \texttt{cause\_of\_injury\_category} (in particular \texttt{IMPACT} and \texttt{LIFTING\_CARRYING}) show substantial predictive power. The complete feature importance ranking is available in the accompanying notebook.

\subsection{Implications for Actuarial Practice}

This case study demonstrates that integrating LLM-derived features into actuarial predictive models can yield meaningful improvements in accuracy. By extracting structured features from unstructured claim descriptions -- body-part categories, cause-of-injury categories, and the number of injured body parts -- and feeding them into a gradient boosting model, we achieve a substantial improvement in claim cost prediction. The ablation study confirms that the body-part and cause-of-injury features each contribute independently, with the combination of all three feature groups producing the best results. Feature importance analysis shows that both traditional and LLM-derived variables meaningfully influence outcomes, enhancing not only predictive accuracy but also interpretability by offering insights into the underlying drivers of high claim costs.

The approach extends beyond claim cost prediction. The same feature-extraction pattern -- applying an LLM to unstructured text, then mapping the results into model-ready categories via a deterministic codebook -- can be adapted for underwriting, fraud detection, and other modelling tasks wherever insurers hold unstructured data that is not yet exploited by their analytical pipelines.

Nonetheless, care must be taken when incorporating LLM outputs. As discussed in Section~\ref{sec:nondeterminism}, LLM completions can exhibit temporal variance; the use of a fixed model version and cached raw outputs in this study ensures reproducibility for the specific results reported here. Broader deployment considerations, including data privacy, model governance, and output stability, are addressed in Section~\ref{sec:challenges_and_considerations}.

%% file: case_study_market_comparison.tex
\section{Case Study 2: GenAI-Driven Market Comparison}
\label{sec:case_study_market_comparison}

\subsection{Introduction}

The second case study explores the application of generative AI to conduct market comparisons, specifically targeting financial and insurance data within the annual reports of insurance companies. The process of extracting and harmonising the data of interest for a comparative analysis is typically challenging due to varying and non-standardised structures in the reports, often requiring labour-intensive manual effort prone to errors and inefficiencies. These publicly available reports are a rich yet complex source of information and comprise diverse formats and content types. We demonstrate how GenAI can streamline the extraction and comparison of key aspects from the 2025 annual reports of three major European insurance groups (AXA, Generali, and Zurich).

Generative AI offers capabilities particularly well-suited for addressing these challenges due to its ability to effectively process unstructured data. This facilitates faster and more accurate extraction of both numerical data (e.g., solvency capital ratios under Solvency II or Swiss Solvency Test (SST), discount rates for insurance contract valuations by duration, contractual service margins) and textual data (e.g., insurer financial strength ratings (IFSR) from agencies, strategies for assessing and
mitigating cyber risk, sensitivity analyses). While the focus here is on financial and insurance data in annual reports, the methodology is highly adaptable and can also be applied, for example, to comparisons of risk reports, sustainability reports, or tariff information (such as services and pricing across different insurance products within a specific market).

To assess the robustness of the extraction pipeline, we benchmark five LLMs -- Claude~Sonnet~4.6, GPT-4.1, GPT-4.1~mini, GPT-5.4, and GPT-5.4~mini -- across three extraction aspects using a shared Retrieval-Augmented Generation (RAG)~\cite{Lewis2020RAG} pipeline, deterministic evaluation against ground-truth reference values, and repeated runs to quantify the stability of the results. High error rates reported on financial question-answering benchmarks, where leading RAG-augmented systems still incorrectly answer or refuse a substantial share of questions~\cite{islam2023financebench}, motivate the rigorous ground-truth validation methodology adopted below.

The complete case study is implemented in a Jupyter notebook available on GitHub\footnote{\url{https://github.com/IAA-AITF/Actuarial-AI-Case-Studies/tree/eaj-v1.0/case-studies/2025/GenAI-driven_market_comparison}}.

\subsection{Approach and Techniques}

\begin{figure}[h!]
\centering
\includegraphics[width=0.825\textwidth]{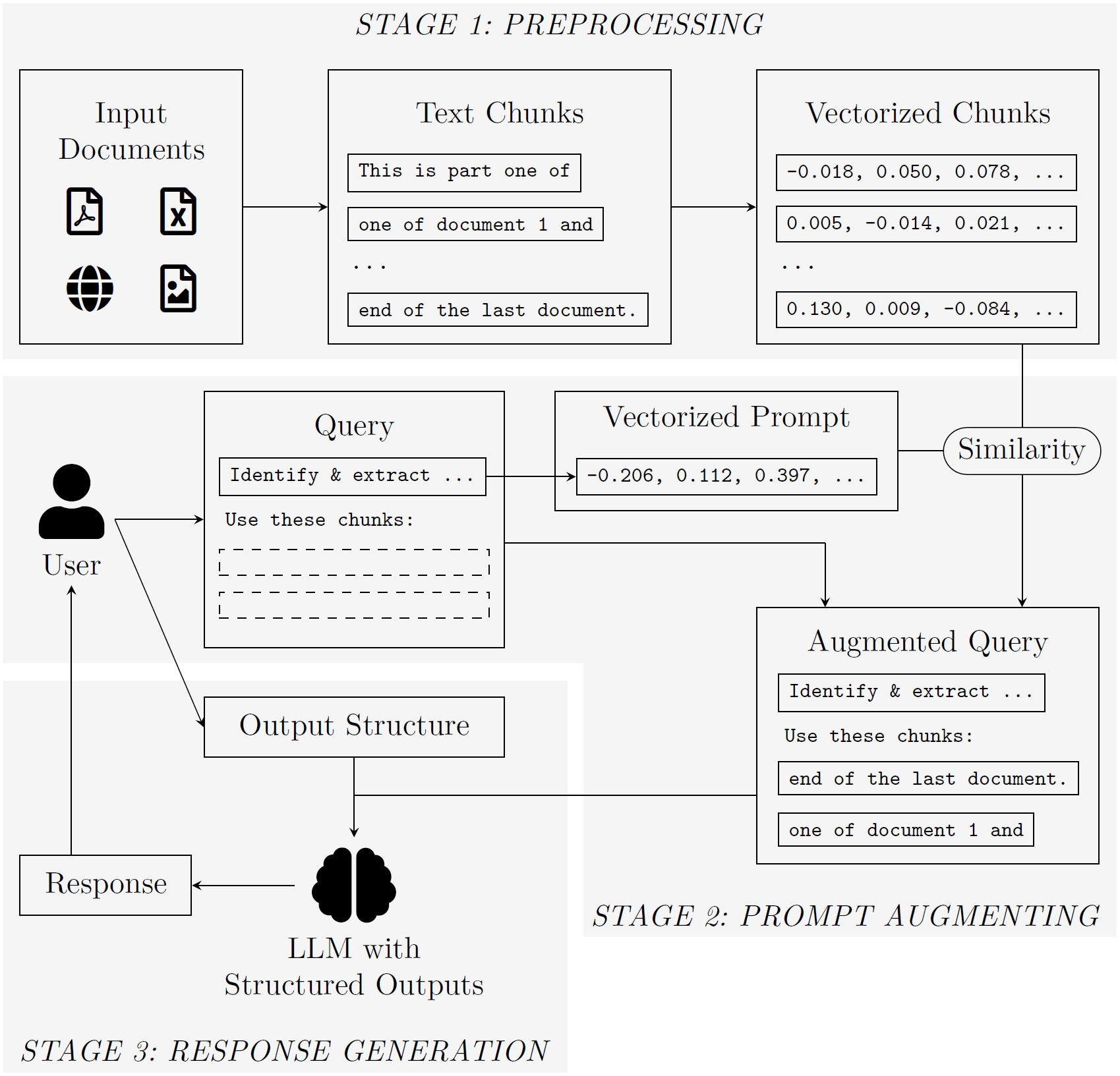}
\vspace{0.3cm}

\caption{The 3-stage approach comprises the stages Preprocessing, Prompt Augmenting, and Response Generation. Preprocessing converts input documents to plain text chunks and corresponding embeddings. Prompt Augmenting retrieves relevant text chunks using vector similarity between the embeddings of the text chunks and the prompt. Response Generation uses an LLM with Structured Outputs to generate a response that precisely adheres to the user-specified format.}
\label{fig:rag_pipeline}
\end{figure}

To demonstrate how to extract relevant information from annual reports in a structured manner, we examine three key aspects: (1) regulatory capital ratios under Solvency II or SST, (2) discount rates for insurance contract valuations by duration, and (3) insurer financial strength ratings. These examples represent distinct data types: a single number with a categorical label, a numerical table of varying length, and a structured list of records. Our approach leverages advanced GenAI techniques, specifically RAG and Structured Outputs. Figure~\ref{fig:rag_pipeline} depicts a schematic illustration of the 3-stage approach, showing how these techniques integrate into the process and how a user's query is transformed into a specified output.

For a fair comparison across models, the entire retrieval pipeline (Stages~1 and~2) is shared and held constant; only the generation model in Stage~3 varies. The five models benchmarked are listed in Table~\ref{tab:model_versions} together with their exact version identifiers. All models are called with temperature~$0.0$ to maximise determinism, and each model--aspect--company combination is executed twenty times to quantify result stability (further run conditions are detailed in Subsection~\ref{sec:case_study_market_comparison_results}).

\begin{table}[h!]
\centering
\caption{LLMs benchmarked in this case study with their provider and exact model version identifier.}
\label{tab:model_versions}
\begin{tabular}{lll}
\toprule
\textbf{Model} & \textbf{Provider} & \textbf{Version Identifier} \\
\midrule
Claude Sonnet 4.6  & Anthropic & \texttt{claude-sonnet-4-6} \\
GPT-4.1            & OpenAI    & \texttt{gpt-4.1-2025-04-14} \\
GPT-4.1 mini       & OpenAI    & \texttt{gpt-4.1-mini-2025-04-14} \\
GPT-5.4            & OpenAI    & \texttt{gpt-5.4-2026-03-05} \\
GPT-5.4 mini       & OpenAI    & \texttt{gpt-5.4-mini-2026-03-17} \\
\bottomrule
\end{tabular}
\end{table}

\subsubsection*{Stage 1: Preprocessing}

The goal of the first stage is to prepare each input document for efficient similarity-based retrieval. This process involves loading and converting the document to plain text, cleansing and chunking the content, and then vectorising it using an embedding model.

Although the following procedure is applicable to various data types -- such as text files, spreadsheets, websites, or images -- we focus on annual reports in PDF format, as this file type is commonly used for such and similar documents, and extract the textual content from each page. Before further processing, text cleansing is performed to improve later retrieval quality by removing irrelevant formatting, adjusting whitespace, and ensuring a consistent text representation -- while preserving the original semantic meaning. Due to the substantial length of annual reports, directly inputting their cleansed content into LLMs is impractical because of context window limitations of these models. To address this, the cleansed texts are segmented into smaller chunks using a simple sliding-window approach: the text is split into chunks of at most 2{,}000~characters with an overlap of 300~characters between consecutive chunks to maintain contextual coherence.

Next, each chunk is transformed into a high-dimensional numerical vector using OpenAI's \texttt{text-embedding-3-large} embedding model, which seeks to map semantically similar content to similar points in the embedding space. The resulting embeddings enable efficient comparisons with the embedded version of the prompt generated in Stage~2. While vector databases are typically used for optimised storage, retrieval, and comparison of embeddings, we chose to simplify the implementation by using working memory instead.

\subsubsection*{Stage 2: Prompt Augmenting}

While the first stage is largely independent of the specific information to be extracted and generally needs to be conducted only once, Stage~2 focuses on identifying the relevant context within the input documents for each of the three key aspects individually.

For each aspect, a specific prompt is formulated that serves a dual purpose: it is first embedded using the same embedding model as in Stage~1 and compared against the chunk vectors to identify relevant context via cosine similarity; the same prompt is then sent to the LLM together with the retrieved chunks to instruct the extraction. The text chunks are ranked based on their similarity scores; the top~10 chunks exceeding a minimum cosine similarity threshold of~0.30 are selected and integrated into the prompt to form the augmented query.

It is important to note that crafting effective prompts is critical for retrieving the most relevant context chunks. The prompts we formulated for the three key aspects are provided below:

\begin{lstlisting}[language=]
# Prompt for solvency capital ratios
prompt_solvency = """
    Extract the group's solvency capital ratio in percentage for 2025, together with the regulatory
    framework (Solvency II or SST).
"""

# Prompt for discount rates
prompt_discount_rates = """
    Extract the discount rates for financial or insurance contract liabilities in 2025, using only 
    currency EUR. For each duration (e.g., 1 year, 5 years, 10 years, 20 years, 40 years, etc.), 
    extract the corresponding discount rate in percentage. Ensure that the data reflects the rates
    as of December 31, 2025. If no specific approach is mentioned, assume non-VFA, unit-linked
    contracts, or liquid products.
"""

# Prompt for insurer financial strength ratings
prompt_ratings = """
    Extract insurer financial strength ratings (IFSR) as a list of entries with rater (e.g., AM 
    Best, Fitch, Moody's, and S&P), rating, and outlook (stable, positive, or negative).
"""
\end{lstlisting}

\subsubsection*{Stage 3: Response Generation}

The final stage involves combining the augmented query from the previous stage with a predefined output structure for each aspect of interest, ensuring that an LLM adheres to this specific format in its response.

To ensure consistent and structured results, we employ Structured Outputs, which guide the LLM to produce responses in a predefined JavaScript Object Notation (JSON) format. By specifying a clear schema -- using JSON or Pydantic syntax -- we explicitly define the expected data types (such as integers, strings, custom data types, or lists of these) and hierarchical levels. The resulting outputs are therefore well-organised, machine-readable, and directly suitable for automated downstream processing, thereby eliminating common inconsistencies that arise in free-form responses.

The output structures we used for the three aspects are shown below in Pydantic syntax:

\begin{lstlisting}[language=]
# Solvency capital ratio schema: percentage and regulatory framework
class SolvencyResult(BaseModel):
    capital_ratio: int          # solvency ratio in %
    regulatory_framework: Literal["Solvency II", "SST"]

# Discount rate for a specific duration
class DiscountRatePoint(BaseModel):
    duration_year: int            # duration in years
    discount_rate_percent: float  # rate in percentage (e.g., 2.47)

# Aggregate discount rates across durations
class DiscountCurveResult(BaseModel):
    currency: Literal["EUR"]
    rates: List[DiscountRatePoint]  # list by duration

# Individual financial strength rating entry
class FinancialStrengthRating(BaseModel):
    rater: str                  # e.g., "S&P Global Ratings", "Moody's"
    rating: str                 # e.g., "AA-", "Aa3", "A+ Superior"
    outlook: Optional[str]      # e.g., "Stable", "Positive"

# Aggregate financial strength ratings
class FinancialStrengthRatingsResult(BaseModel):
    ratings: List[FinancialStrengthRating]
\end{lstlisting}

To enable the multi-model benchmark, we use the LangChain framework\footnote{\url{https://www.langchain.com/}}, which provides a unified interface for invoking different LLM providers (e.g., OpenAI, Anthropic) with Structured Outputs through a common API. This allows the same extraction logic to be applied identically across all five models.

\subsubsection*{Customisation and Advanced RAG Variants}

The 3-stage approach outlined above can be customised in various ways to enhance precision and adapt to specific requirements. One method involves employing fine-tuned LLMs trained on domain-specific data, as illustrated in the case study in Section~\ref{sec:case_study_vision_fine_tuning}. Additionally, the approach can be refined by modifying the extraction, cleansing, and chunking processes of input documents to better align with the nature and complexity of the data. Customising the prompt and optionally integrating metadata such as keywords can further improve the retrieval of relevant information. Adjustments to parameters like the number of context chunks considered relevant and the threshold for their inclusion ensure that only the most pertinent data is utilised to augment the query.

Furthermore, several advanced RAG concepts hold potential for further enhancing the overall process (see also Raieli and Iuculano~\cite{raieli2025building} for a comprehensive treatment of RAG pipelines, knowledge graphs, and agentic architectures):

\begin{itemize}
\item \textbf{GraphRAG} \cite{edge2025localglobalgraphrag}: Enhances the RAG system by incorporating a knowledge graph that explicitly models entities and their relationships. In particular, multi-hop reasoning across related entities can resolve ambiguities that single-step retrieval may miss.
\item \textbf{Agentic RAG} \cite{singh2025agenticrag}: Introduces AI agents (compare Section~\ref{sec:multi_agent_system}) into the RAG pipeline, enabling dynamic decision-making and tool selection during the retrieval process. These agents can determine the necessity of external information, select appropriate data sources, and iteratively refine retrieval strategies based on the complexity of the user query.
\end{itemize}

\subsection{Results}
\label{sec:case_study_market_comparison_results}

\subsubsection*{Run Conditions and Validation}

All experiments use temperature~$0.0$, which instructs each provider to return the most likely token at every decoding step, aiming to maximise determinism. Minor run-to-run variations may still occur due to floating-point non-determinism in GPU computations and provider-side infrastructure changes. No additional seed parameter is used, as not all providers expose one. Each model--aspect--company combination is executed twenty times, yielding $5 \times 3 \times 3 \times 20 = 900$ individual extractions across the benchmark. The embedding model (\texttt{text-embedding-3-large}) and all retrieval parameters are held constant throughout. All models are referenced by pinned version identifiers (Table~\ref{tab:model_versions}) to ensure reproducibility; broader considerations on non-determinism and model drift are discussed in Section~\ref{sec:nondeterminism}.

To rigorously assess extraction quality, we define ground-truth reference values for each of the nine company--aspect combinations (three companies~$\times$ three aspects). These reference values were manually extracted by the authors from the original 2025 annual reports. In total, the ground truth comprises 45~individual fields: 3~solvency ratios (each with an associated regulatory framework label), 33~discount-rate points (10~for AXA, 18~for Generali, and 5~for Zurich), and 9~financial strength rating entries (3~per company).

\subsubsection*{Ground-Truth Reference Values and Evaluation}

Tables~\ref{tab:gold_solvency},~\ref{tab:gold_discount_rates}, and~\ref{tab:gold_ratings} present the ground-truth reference values used for evaluation. Since the underlying annual reports are publicly available, reporting these values enables independent verification.

\begin{table}[h!]
\centering
\caption{Ground-truth solvency capital ratios and regulatory frameworks for 2025.}
\label{tab:gold_solvency}
\begin{tabular}{lrl}
\toprule
\textbf{Company} & \textbf{Capital Ratio} & \textbf{Regulatory Framework} \\
\midrule
AXA      & 224\% & Solvency II \\
Generali & 219\% & Solvency II \\
Zurich   & 259\% & SST \\
\bottomrule
\end{tabular}
\end{table}

\begin{table}[h!]
\centering
\caption{Ground-truth EUR discount rates (\%) for insurance contract liabilities as of 31 December 2025. A dash indicates that the company does not report a rate for that duration.}
\label{tab:gold_discount_rates}
\begin{tabular}{rccc}
\toprule
\textbf{Duration (years)} & \textbf{AXA} & \textbf{Generali} & \textbf{Zurich} \\
\midrule
1  & 2.40 & 2.22 & 2.08 \\
2  & 2.50 & 2.30 & --   \\
3  & 2.60 & 2.42 & --   \\
4  & --   & 2.53 & --   \\
5  & 2.80 & 2.62 & 2.48 \\
6  & --   & 2.70 & --   \\
7  & 3.00 & 2.79 & --   \\
8  & --   & 2.86 & --   \\
9  & --   & 2.93 & --   \\
10 & 3.20 & 3.00 & 2.86 \\
15 & 3.40 & 3.25 & --   \\
20 & 3.50 & 3.35 & 3.21 \\
25 & 3.50 & 3.39 & --   \\
30 & 3.40 & 3.41 & --   \\
35 & --   & 3.41 & --   \\
40 & --   & 3.40 & 3.27 \\
45 & --   & 3.40 & --   \\
50 & --   & 3.39 & --   \\
\bottomrule
\end{tabular}
\end{table}

\begin{table}[h!]
\centering
\caption{Ground-truth insurer financial strength ratings (IFSR) for 2025.}
\label{tab:gold_ratings}
\begin{tabular}{llll}
\toprule
\textbf{Company} & \textbf{Agency} & \textbf{Rating} & \textbf{Outlook} \\
\midrule
AXA      & S\&P     & AA-         & Positive \\
AXA      & Moody's  & Aa2           & Stable \\
AXA      & AM Best  & A+ Superior   & Stable \\
\midrule
Generali & Moody's  & A2            & Stable \\
Generali & Fitch    & AA-         & Stable \\
Generali & AM Best  & A+            & Stable \\
\midrule
Zurich   & S\&P     & AA            & Stable \\
Zurich   & Moody's  & Aa2           & Stable \\
Zurich   & AM Best  & A+ Superior   & Stable \\
\bottomrule
\end{tabular}
\end{table}

Each extraction is evaluated using exact accuracy: the result is scored as either~$1$ (pass) or~$0$ (fail), with no partial credit. The aspect-specific pass criteria are:

\begin{itemize}
\item \textbf{Solvency ratios}: The capital ratio (integer percentage) and the regulatory framework label must both match the ground truth exactly.
\item \textbf{Discount rates}: For every duration in the ground truth, the extracted rate must match the ground-truth value exactly at the precision reported in the annual reports (two decimal places). Rounding to this precision merely removes representation artefacts (e.g., $2.4$ versus $2.40$) and applies no tolerance band. All durations must be present for the extraction to pass.
\item \textbf{Financial strength ratings}: Normalised set comparison with alias-aware matching. Rater names are mapped to canonical forms (e.g., ``S\&P Global Ratings'' and ``Standard \& Poor's'' both map to ``S\&P''), ratings are uppercased with parenthetical variations removed, and outlooks are title-cased. All ground-truth entries must be matched for the extraction to pass.
\end{itemize}

We deliberately adopt this exact, binary pass/fail criterion rather than a tolerance band or partial credit. In a market-comparison setting, the extracted figures feed directly into regulatory and competitive analyses, where a solvency ratio, a discount curve, or an insurer rating is only operationally useful if it reproduces the published value exactly; an approximate or partially correct extraction would propagate silently into the downstream comparison. A strict per-field criterion therefore mirrors the acceptance decision an actuary would apply in practice and sets a necessarily demanding bar for the benchmark.

\subsubsection*{Benchmark Results}

Table~\ref{tab:benchmark_pass_rates} summarises the pass rates across the five models and three extraction aspects. Each cell reports the percentage of runs (out of twenty per company) for which the extraction fully matched the ground truth, averaged across the three companies.

\begin{table}[h!]
\centering
\caption{Pass rates (\%) by model and extraction aspect, averaged across twenty runs and three insurance groups (AXA, Generali, Zurich).}
\label{tab:benchmark_pass_rates}
\begin{tabular}{lccc}
\toprule
\textbf{Model} & \textbf{Solvency Ratios} & \textbf{Discount Rates} & \textbf{Fin.\ Strength Ratings} \\
\midrule
Claude Sonnet 4.6  & 100.0 & 100.0 & 100.0 \\
GPT-4.1            & 100.0 & 100.0 & 100.0 \\
GPT-4.1 mini       & 100.0 & 100.0 & 66.7 \\
GPT-5.4            & 100.0 & 100.0 & 100.0 \\
GPT-5.4 mini       & 100.0 & 100.0 & 71.7 \\
\bottomrule
\end{tabular}
\end{table}

GPT-4.1, GPT-5.4, and Claude~Sonnet~4.6 achieve perfect pass rates across all nine company--aspect combinations over twenty runs each. The two smaller models are weaker only on financial strength ratings, where GPT-5.4~mini reaches 71.7\% and GPT-4.1~mini 66.7\%. Solvency ratio and discount rate extraction prove fully reliable across all five models, including under the exact-match criterion described above.

\subsubsection*{Failure Modes}

Across the benchmark, we observe several characteristic failure modes:
\begin{itemize}
\item \textbf{Solvency ratios.} This is the most reliable aspect; no failures were observed across any model or run.
\item \textbf{Discount rates.} Potential failures include extraction of rates for a different currency or valuation date, rounding errors, and missing entries for specific durations. Discount-rate tables that combine multiple currencies and valuation dates in a single table increase the risk of misattribution. None of these failures occurred in the present benchmark.
\item \textbf{Financial strength ratings.} The most common error is the inclusion of issuer credit ratings or senior debt ratings instead of the requested IFSR -- a distinction that requires careful prompt engineering. Entity-level ambiguity poses a further challenge: AXA's annual report separates ratings for AXA~SA (the holding company) from those for its primary insurance subsidiaries, leading smaller models to select inconsistently between the two.
\end{itemize}
These observations align with the taxonomy of RAG errors proposed by Leung et al.~\cite{leung2026rag_errors} and the broader hallucination taxonomy of Huang et al.~\cite{huang2025hallucination}; both works distinguish retrieval failures (insufficient or irrelevant context) from generation failures (incorrect reasoning over correct context). The ratings failures in our benchmark are generation-stage errors: the relevant context is retrieved correctly, but the model inconsistently resolves entity-level ambiguity. Retrieval failures are typically addressed by adjusting the prompt or retrieval parameters, while generation failures may require more explicit instructions or a more capable model.

\subsection{Implications for Actuarial Practice}

The application of advanced generative AI techniques, particularly RAG and Structured Outputs, demonstrates the potential of LLMs to streamline the extraction and comparison of complex data across large, unstructured documents. Using annual reports of insurance companies as an example, this approach shows how actuaries can obtain structured, comparable insights with greater efficiency, consistency, and reduced risk of errors. The RAG framework proves its value by addressing the inherent limitations of LLM context windows, allowing actuaries to access precise, focused information from lengthy and technical reports. Structured Outputs further ensure that extracted insights are consistently formatted, which is essential for their seamless integration into downstream analysis pipelines, reducing variability and supporting reliable processing across diverse data sources and reporting standards.

The multi-model benchmarking reveals that retrieval quality -- governed by chunking strategy, embedding model, and prompt design -- matters as much as the choice of generation model; improvements to the retrieval stage often yield larger gains than switching to a more capable LLM, which highlights the importance of investing in document preprocessing and prompt engineering. The financial strength ratings aspect, in particular, illustrates the importance of prompt specificity: when multiple related but distinct data items coexist in a document, explicit instructions are needed to ensure the correct items are extracted. This underscores the continued need for domain expertise in prompt engineering, and actuarial expertise remains indispensable at multiple stages of the process, including refining prompts, identifying ground-truth reference values, validating results, and adapting output structures to specific use cases.

The current approach is particularly effective for the selected aspects of annual reports -- solvency capital ratios, discount rates, and insurer financial strength ratings -- which are well-defined and structured. Applying the same methods to less structured or more complex areas often requires iterative refinement, careful oversight, and tailored prompt engineering to achieve similarly robust results. When the desired information is not present in a document, strategies must be in place to handle such cases explicitly: for example, instructing the LLM to indicate when no appropriate context is found helps prevent misleading outputs and ensures that analysts are aware of data limitations. While the entire data extraction and comparison process could, in theory, be fully automated, maintaining human oversight is strongly recommended; providing actuaries with interim outputs -- such as the extracted plain text, retrieved chunks for review, or comparisons with historical results -- supports validation, ensures accuracy, and reinforces trust in AI-driven analyses.

%% file: case_study_vision_fine_tuning.tex
\section{Case Study 3: Car Damage Classification and Localisation with Fine-Tuned Vision-Enabled LLMs}
\label{sec:case_study_vision_fine_tuning}

\subsection{Introduction}

This case study explores how large language models can improve both the classification and contextual understanding of car damage from images -- an important task in automotive insurance, particularly for claims processing and risk assessment. Traditional computer vision methods, such as convolutional neural networks (CNNs), have demonstrated strong performance in static image classification~\cite{van2022convolutional, PerezZarate2024CarDamage, hasan2025vehicledamage}. However, these models often struggle to additionally incorporate contextual information that is valuable for insurance applications, such as precisely localising damage, evaluating its severity, and accounting for external factors such as lighting and weather conditions at the time of capture.

To address these limitations, we employ OpenAI's GPT-4o (version \texttt{gpt-4o-2024-08-06}), a vision-enabled LLM that integrates image recognition with natural language understanding. By fine-tuning this model on a domain-specific dataset of labelled car damage images, we achieve classification performance that is comparable to traditional models while also providing richer contextual insights. This enhanced capability allows the model to distinguish, for example, between minor glass damage on a side window and a fully shattered windshield.

Beyond car damage analysis, this approach demonstrates broad applicability across various visual tasks in insurance. Its flexibility extends to medical image analysis, fraud detection in claims and invoices, and roof damage assessment in household and commercial property insurance, among others. Notably, the INS-MMbench dataset~\cite{lin2024insmmbench} provides a diverse collection of images covering these and other insurance-related tasks.

The full case study presented below has been implemented as a Jupyter notebook, which is available on GitHub\footnote{\url{https://github.com/IAA-AITF/Actuarial-AI-Case-Studies/tree/eaj-v1.0/case-studies/2025/car_damage_classification_and_localization}}.

\subsection{Approach and Techniques}

In this case study, we use a dataset on car damages from a Kaggle competition\footnote{\url{https://www.kaggle.com/datasets/imnandini/analytics-vidya-ripik-ai-hackfest}}. The dataset consists of thousands of car images, each labelled with one of six damage types: crack, scratch, tire flat, dent, glass shatter, or lamp broken.  Table~\ref{tab:damage_examples_locations} shows three example images illustrating different damage types.

Our primary objective is to develop a supervised learning model that predicts the damage class of a given image, while our secondary goal is to extract the precise location of the damage. To this end, we compare three models:
\begin{itemize}
    \item A classical convolutional neural network
    \item The standard, off-the-shelf version of OpenAI's GPT-4o
    \item A version of GPT-4o fine-tuned on a subset of our dataset
\end{itemize}
While all three models can perform the primary classification task, the CNN lacks the ability to capture contextual information relevant to the second objective. We note that the goal of this case study is not to maximise the predictive performance of any individual model, but rather to demonstrate how fine-tuning can improve LLM-based classification, to illustrate the contextual capabilities of vision-enabled LLMs, and to compare the LLM-based and CNN-based approaches.
For the supervised learning task, we take a sample of 1,500 images and apply a conventional data split, dividing the dataset into training (60\%), validation (20\%), and test (20\%) subsets.
The CNN is trained and validated on the respective subsets, and its performance is evaluated on the unseen test data. Accuracy and the weighted F1 score (accounting for the number of true instances per label) are used as evaluation metrics, both of which are standard for multiclass classification.

In contrast, OpenAI's GPT-4o -- where the `P' in the acronym aptly stands for \textit{pre-trained} -- has already been trained on a large and diverse set of images, allowing it to leverage its built-in vision capabilities to directly classify images without further training. For each test image, we combine a system prompt (which provides instructions to the LLM) with the image's base64-encoded textual representation and use the Structured Outputs concept (cf. the case study in Section~\ref{sec:case_study_market_comparison}) to ensure that the model classifies the image into one of the predefined damage types. The following system prompt is employed:

\begin{lstlisting}[language=]
system_prompt = """
    You will classify images of cars with visible damages into one of six specific classes. Your
    task is to examine the image provided, identify the type of damage, and return the correct
    damage type as one of the predefined classes. The six damage classes are as follows:

    1. crack
    2. scratch
    3. tire flat
    4. dent
    5. glass shatter
    6. lamp broken

    # Steps

    1. **Analyze the Image:**
       - Carefully inspect the image provided, focusing on visible damage to determine its type.
       - Pay attention to details such as patterns, locations, and characteristics of the damage
         to accurately classify it.

    2. **Identify the Correct Class:**
       - Match the observed damage to one of the six predefined classes listed above.
       - Avoid ambiguous classifications and ensure the answer aligns precisely with the provided
         options.

    3. **Output the Classification:**
       - Provide your answer in a concise format with **only the name of the damage class** (e.g.,
         "scratch").
       - Do not add any additional text, reasoning, or explanations. The response must be exactly
         one of the six class names.

    # Notes
    - If multiple damage types appear equally prominent in the image, select the one that seems
      most severe or predominant.
    - Ensure every response falls strictly within the six defined classes. Avoid assumptions or
      interpretations beyond the scope of these categories.

    # Example Format
    For an input image with a visible scratch:
    - Correct Output: "scratch"

    For an input image showing a shattered glass window:
    - Correct Output: "glass shatter"

    Adhere to this strict output format and guidelines in every classification.
"""
\end{lstlisting}

Next, we fine-tune GPT-4o using our training and validation datasets in order to improve its classification performance. For this purpose, we utilise OpenAI's fine-tuning platform\footnote{For more (technical) details on fine-tuning OpenAI's LLMs, see \url{https://platform.openai.com/docs/guides/fine-tuning}.} providing it with a set of input-output pairs. Each input consists of the previously used system prompt combined with the image's base64-encoded textual representation, while the output is the corresponding label. Notably, other multimodal LLMs, such as Google's Gemini or Meta's Llama, also support image fine-tuning. Once fine-tuning is complete -- which may take several hours -- the enhanced model is evaluated on the test set images using the same system prompt as before.

For the secondary objective, we employ the fine-tuned GPT-4o and adjust the system prompt to enable predictions of both the damage type and, when identifiable, the damage location. Additionally, we refine the Structured Output format so that the model outputs one of the six classes and, optionally, the damage location. The modified prompt can be found in the accompanying Jupyter notebook. 

For simplicity, we have focused solely on damage localisation. Other aspects of visual context that could be extracted include damage severity, external factors such as lighting and weather conditions, and additional details like vehicle make and licence plate information.

\subsection{Results}

Table~\ref{tab:vision_comparison} presents a comparative analysis of the CNN, the non-fine-tuned GPT-4o, and the fine-tuned GPT-4o, using accuracy and weighted F1 score as evaluation metrics on the test data. The results show that fine-tuning GPT-4o improves its performance, with the fine-tuned model achieving 0.880 for both evaluation metrics, outperforming the non-fine-tuned version (0.823 accuracy and 0.825 weighted F1 score). Moreover, the fine-tuned GPT-4o achieves performance comparable to the convolutional neural network (0.837 accuracy and 0.835 weighted F1 score).

\begin{table}[h!]
\centering
\caption{Comparison of multiclass classification performance on the test data of the car damage dataset, evaluated using accuracy and weighted F1 score. The table compares three models: a convolutional neural network, the non-fine-tuned GPT-4o, and the fine-tuned GPT-4o. For both metrics, higher values indicate better performance.}
\label{tab:vision_comparison}
\begin{tabular}{lcc}
\toprule
\textbf{Prediction Model} & \textbf{Accuracy $(\uparrow)$} & \textbf{Weighted F1 Score $(\uparrow)$} \\
\midrule
Convolutional Neural Network & 0.837 & 0.835  \\
Non-Fine-Tuned GPT-4o & 0.823 & 0.825 \\
Fine-Tuned GPT-4o & 0.880 & 0.880 \\
\bottomrule
\end{tabular}
\end{table}

To demonstrate the model's contextual capabilities, we use the fine-tuned GPT-4o model to generate both the predicted damage type and the damage location of the three example images in Table~\ref{tab:damage_examples_locations}. As exemplarily shown in this table, the model correctly identifies the damage locations -- for instance, indicating that the glass shatter is on the windshield (left image) and the dent is on the rear bumper (right image) -- while in one case (middle image, tire flat), the specific tire could not be determined. Note that since the dataset does not include damage locations, the correctness of the predicted damage locations has to be verified manually.

\begin{table}[h!]
    \setlength{\tabcolsep}{2pt}
    \centering
    \caption{Comparison of actual and predicted damage types and locations for selected car damage images. Each column shows an input image, the true damage type, and the damage type and (if identifiable) damage location predicted by the fine-tuned GPT-4o model.}
    \label{tab:damage_examples_locations}
    \begin{tabular}{l c c c}
        \toprule
        &
        \includegraphics[width=0.2\linewidth,keepaspectratio]{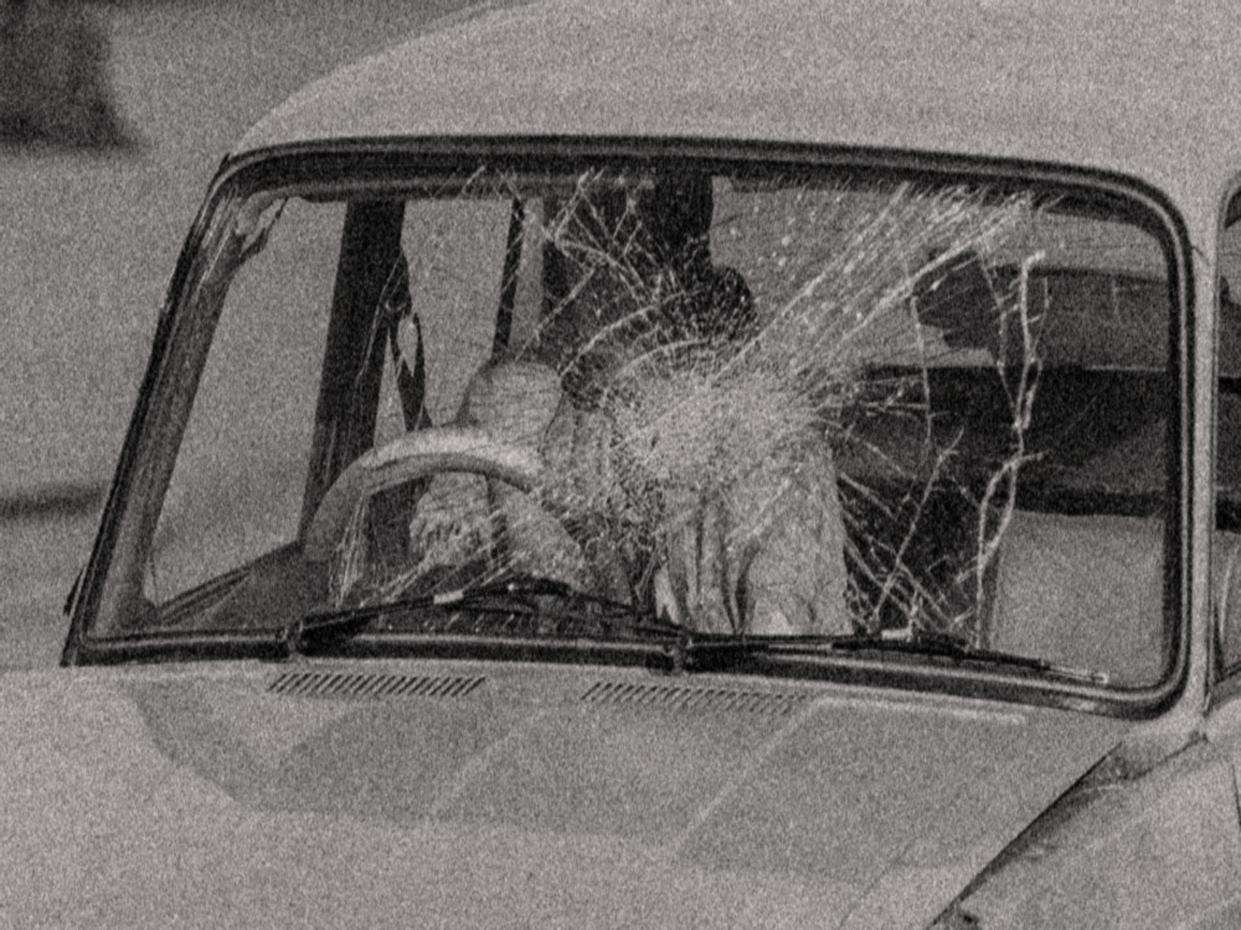} &
        \includegraphics[width=0.2\linewidth,keepaspectratio]{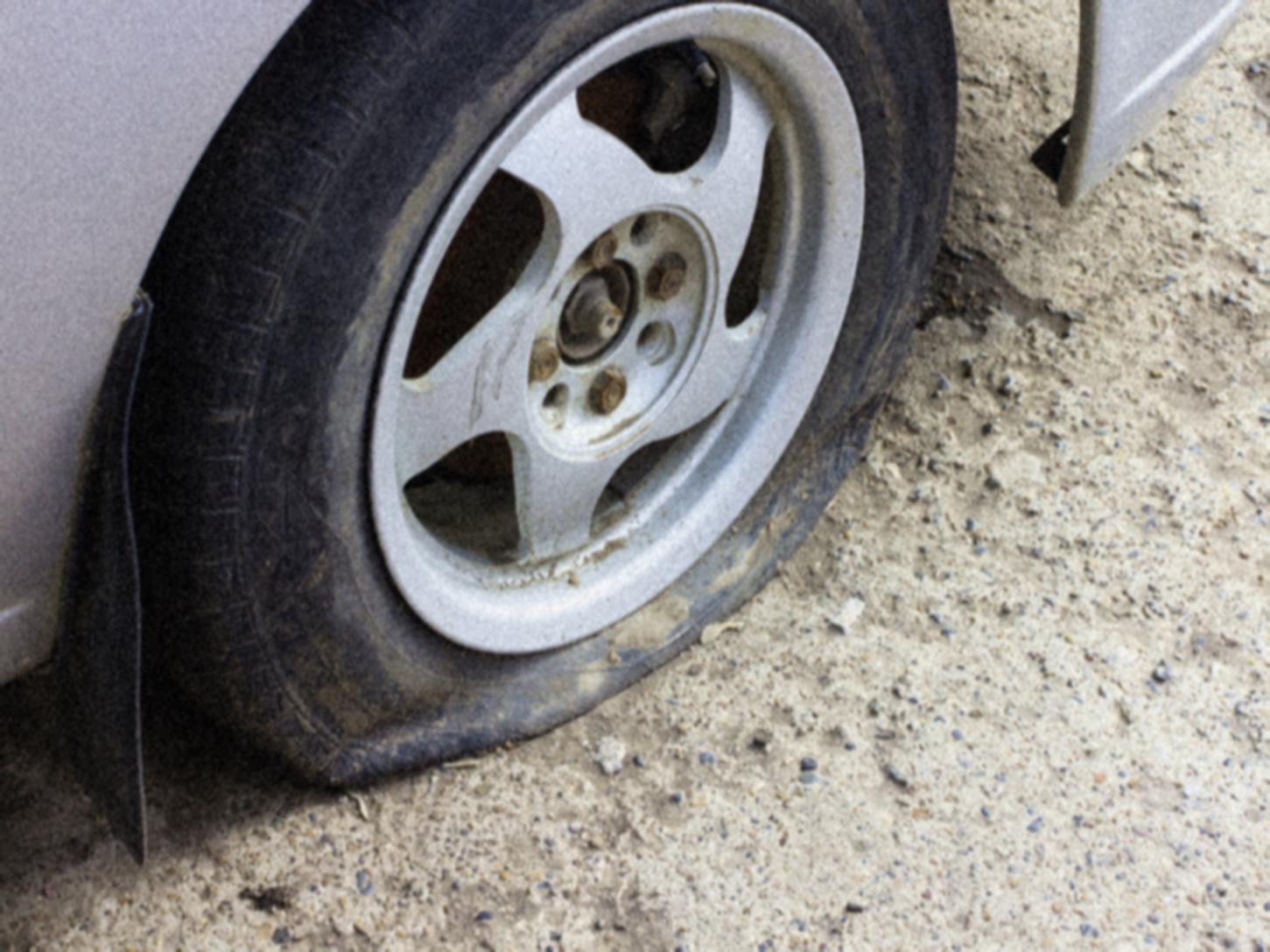} &
        \includegraphics[width=0.2\linewidth,keepaspectratio]{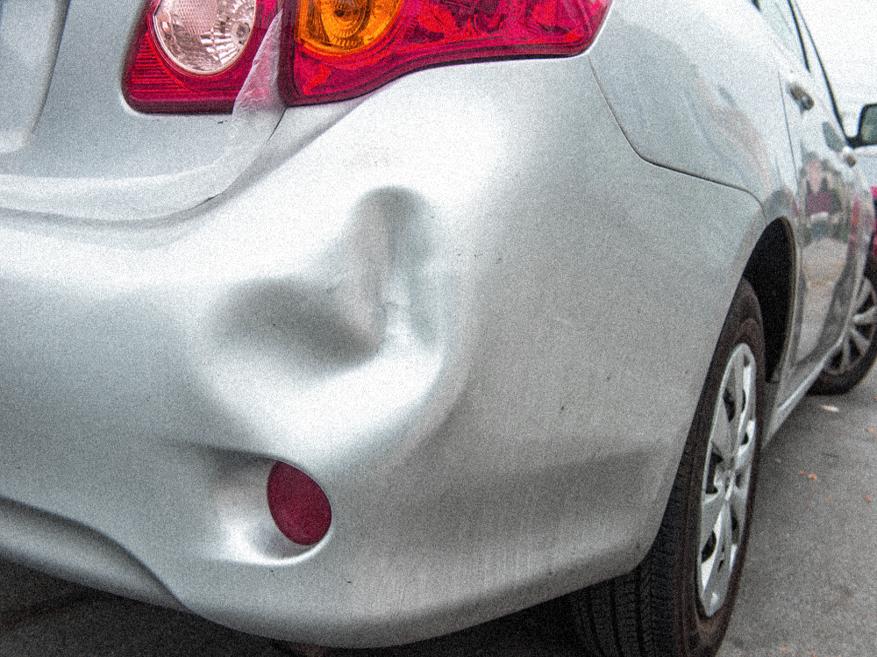} \\
        \midrule
        \textbf{Actual damage type} & glass shatter & tire flat & dent \\
        \textbf{Predicted damage type} & glass shatter & tire flat & dent \\
        \textbf{Predicted damage location} & windshield & --- & rear bumper \\
        \bottomrule
    \end{tabular}
\end{table}

\subsection{Implications for Actuarial Practice}

The application of advanced generative AI techniques in car damage classification demonstrates the potential of vision-enabled LLMs to improve classification performance and extract richer contextual insights, with capabilities that extend well beyond this domain. Fine-tuning a vision-enabled LLM on domain-specific data can improve classification performance -- as demonstrated in our experiments, where off-the-shelf models provide a solid baseline -- and such benefits typically apply not only to visual tasks but also to textual applications.

Beyond classification, vision-enabled LLMs can capture contextual details and even perform optical character recognition (OCR). While traditional CNNs are effective at object recognition, they struggle with contextual analysis; LLMs, by contrast, provide a more holistic assessment of image information and can offer preliminary estimates of damage severity and repair costs. The use of Structured Outputs further ensures that model predictions adhere to predefined categories, eliminating inconsistencies and guaranteeing that each prediction corresponds to one of the prespecified damage classes.

A further advantage of LLMs is their ease of use compared with CNNs, which require expertise in model architecture and optimisation. With fine-tuning, LLMs can achieve high-quality results with comparatively less engineering effort for model development, enabling actuaries to integrate AI-driven visual analysis into their workflows without specialised deep learning knowledge. On the downside, the fine-tuning process may take considerably longer than training a CNN, and moving from a working prototype to a production-ready system requires substantial additional investment in data governance, monitoring, and human oversight -- considerations that are discussed further in Section~\ref{sec:challenges_and_considerations} and should not be underestimated.

We also note that our evaluation is limited to a single CNN baseline on a sample of 1{,}500 images and does not systematically assess robustness across capture conditions, inference cost and latency, or the privacy implications of transmitting claim images to a third-party API -- dimensions that warrant careful attention before any production deployment.

However, with the growing image generation capabilities of LLMs, there is a risk of misuse for fraud. For example, a car without damage could be altered to show realistic damage at a specific location and severity, potentially bypassing traditional verification methods -- without the need for professional image manipulation skills. The risks of synthetic media for insurance fraud, including detection and forensic countermeasures, provenance verification, and escalation protocols, are discussed in detail in Section~\ref{sec:challenges_and_considerations}; see also~\cite{mitra2024} for a broader review of deepfake risks.

%% file: case_study_multi_agent_system.tex
\section{Case Study 4: Actuarial Legacy Code Migration Multi-Agent System}
\label{sec:multi_agent_system}

\subsection{Introduction}

At the Consumer Electronics Show (CES) 2025, NVIDIA CEO Jensen Huang proclaimed the era of ``agentic AI'', reflecting a growing interest in intelligent systems capable of executing complex tasks with limited human involvement. Central to this paradigm are AI agents -- autonomous software entities that perceive their environment, make decisions, and act to achieve specific goals. Unlike general-purpose LLMs, which are mainly used to generate text based on input prompts, AI agents can plan, reason, and interact with external tools such as code interpreters, databases, and web search engines, enabling them to complete more intricate and context-aware tasks~\cite{pan2024multiagent, liu2026llmseagents}.

Building on this foundation, multi-agent systems (MAS) consist of multiple AI agents operating collaboratively within a shared environment to solve problems that are beyond the scope of any individual agent~\cite{wu2023autogen, hong2024metagpt}. Each agent in a MAS is designed with specialised capabilities and can communicate and coordinate with other agents to achieve collective objectives. This architecture allows complex tasks to be decomposed into manageable subtasks, with each agent contributing its domain expertise, resulting in more efficient and scalable solutions.

Legacy code migration -- the translation of existing software from one programming language to another -- represents a particularly compelling application of MAS in the actuarial domain. Many insurance companies maintain critical legacy code -- such as actuarial models, reserving pipelines, or policy administration systems -- written in programming languages like R, SAS, COBOL, or VBA, and face growing pressure to consolidate these into modern ecosystems -- most commonly Python-based -- for maintainability, scalability, and integration with broader data science and AI workflows. Manual migration is time-consuming, error-prone, and requires expertise in both the source and target languages as well as in the underlying actuarial domain. While deterministic parser-based transpilers offer one solution, they may struggle with language-specific idioms, implicit conventions, and domain-specific logic that cannot be captured by syntactic rules alone.

The case study presented in the following demonstrates a practical implementation of a MAS composed of five specialised agents that automates the migration of actuarial R code to Python with ground-truth validation. Specifically, our MAS consists of: (1) an \emph{R analysis agent} that examines the source code and analyses the computational logic; (2) a \emph{translation agent} that converts the R code to equivalent Python code; (3) a \emph{compilation agent} that validates syntax and executes the translated code; (4) a \emph{test runner agent} that runs a pre-written, R-verified test suite; and (5) a \emph{report agent} that produces a structured migration report. The system features automatic retry loops: when compilation or tests fail, the translation agent receives the error details and iteratively refines its output; once a prespecified maximum number of retries is exceeded, the pipeline escalates the failure rather than continuing to retry. This modular, feedback-driven design illustrates how complex migration tasks can be decomposed, validated, and documented through agentic collaboration.

The architecture presented here is not limited to R-to-Python migration. The same multi-agent design generalises to other migration scenarios -- such as SAS-to-Python or VBA-to-Python -- by adapting agent prompts and tools. More broadly, the MAS paradigm is applicable wherever actuarial workflows can be decomposed into discrete, coordinated subtasks -- for example, automated monitoring and alerting pipelines, regulatory reporting workflows, or extensions of the market comparison framework described in Section~\ref{sec:case_study_market_comparison}.

Like the other case studies in this article, the full implementation is provided in a Jupyter notebook on GitHub\footnote{\url{https://github.com/IAA-AITF/Actuarial-AI-Case-Studies/tree/eaj-v1.0/case-studies/2026/actuarial_legacy_code_migration_multi-agent_system}}.

\subsection{Approach and Techniques}

We build a legacy code migration multi-agent system powered by LLMs from OpenAI and orchestrated via the LangGraph framework\footnote{\url{https://github.com/langchain-ai/langgraph}}, which supports stateful, graph-based agent workflows. Unlike the supervisor-based pattern commonly used in MAS~\cite{wu2023autogen}, our system employs a hardcoded sequential workflow with conditional retry loops: the agent execution order is determined by explicit graph edges, not by LLM routing decisions.

Each agent is powered by a distinct LLM variant selected for its task-specific strengths -- GPT-5.4 for complex code analysis and translation tasks, and GPT-5.4 mini for simpler validation and reporting tasks. The overall architecture of the legacy code migration MAS is illustrated in Figure~\ref{fig:migration_mas_architecture}. A detailed description of each agent's functionality follows below.

\begin{figure}[h!]
\centering
\begin{tikzpicture}[
    >=Stealth,
    node distance=0.55cm,
    agent/.style={
        rectangle, draw=black, thick,
        align=center, text width=3.8cm, minimum height=0.95cm,
        fill=black!5, font=\small, inner sep=3pt
    },
    terminal/.style={
        ellipse, draw=black, thick,
        align=center, minimum width=1.6cm, minimum height=0.5cm,
        fill=none, font=\small\bfseries
    },
    flow/.style={->, thick},
    retry/.style={->, thick, dashed, gray!50!black},
    flab/.style={font=\scriptsize\itshape, inner sep=2pt}
]

\node[terminal] (start) {start};
\node[agent, below=of start] (analysis) {\textbf{r\_analysis\_agent}\\[1pt]\scriptsize GPT-5.4};
\node[agent, below=of analysis] (translate) {\textbf{translation\_agent}\\[1pt]\scriptsize GPT-5.4};
\node[agent, below=of translate] (compile) {\textbf{compilation\_agent}\\[1pt]\scriptsize GPT-5.4 mini};
\node[agent, below=of compile] (test) {\textbf{test\_runner\_agent}\\[1pt]\scriptsize GPT-5.4 mini};
\node[agent, below=of test] (report) {\textbf{report\_agent}\\[1pt]\scriptsize GPT-5.4 mini};
\node[terminal, below=of report] (end) {end};

\draw[flow] (start) -- (analysis);
\draw[flow] (analysis) -- (translate);
\draw[flow] (translate) -- (compile);
\draw[flow] (compile) -- (test);
\draw[flow] (test) -- (report);
\draw[flow] (report) -- (end);

\draw[retry] (compile.west) to[out=180, in=180, looseness=1.8]
    node[flab, left, midway, xshift=-2pt]{compile fail} (translate.west);
\draw[retry] (test.east) to[out=0, in=0, looseness=1.8]
    node[flab, right, midway, xshift=2pt]{test fail} (translate.east);
\end{tikzpicture}
\vspace{0.3cm}
\caption{Architecture of the legacy code migration multi-agent system. Five specialised agents process R source code through analysis, translation, compilation, testing, and reporting stages. Conditional retry loops (dashed) route compilation or test failures back to the translation agent for iterative refinement; once a prespecified maximum number of retries (set to 5 in this case study) is exceeded, the pipeline escalates to the report agent.}
\label{fig:migration_mas_architecture}
\end{figure}

\subsubsection*{R Analysis Agent}

The \texttt{r\_analysis\_agent} is powered by GPT-5.4 and serves as the entry point of the pipeline. It reads the R source file and any associated data files (e.g., CSV files, SQLite databases), analyses the code's structure and computational logic, and returns a structured JSON summary of its findings -- including identified functions, data files, libraries used, and key outputs. It can execute R snippets via a dedicated \texttt{run\_r\_code} tool to understand intermediate or final computations.

Importantly, the R analysis agent does \emph{not} generate the test suite. The tests are pre-written, manually audited artefacts prepared in advance of each pipeline run. This design ensures that the evaluation is identical across runs and independent of the translation process. The test suites cover two aspects: the structural properties of the translated code, and the numerical values it produces, checked against ground truth derived from the original R code. The ground-truth values are held separately and deliberately kept outside the translation agent's context, so that it cannot succeed by simply injecting the answers.

The following code snippet illustrates the initialisation of the \texttt{r\_analysis\_agent}, showing the model selection, available tools, and an abbreviated version of the system prompt that governs its behaviour.

\begin{lstlisting}[language=]
# -- Agent 1: R Analysis Agent --
r_analysis_agent = create_agent(
    model="openai:gpt-5.4-2026-03-05",
    tools=[read_file, read_csv_preview, run_r_code],
    name="r_analysis_agent",
    system_prompt=(
        "You are an expert R programmer and actuarial scientist. Your job is to:\n"
        "1. Read the provided R source file using the read_file tool.\n"
        "2. Read any associated data files (CSV) using read_csv_preview to understand inputs.\n"
        "3. If needed, use run_r_code to execute R snippets and understand intermediate values.\n"
        "4. Analyze what the R code does: identify functions, computations, data flow.\n\n"
        "IMPORTANT: Pre-written test files and expected values have already been copied to\n"
        "output/tests/. Do NOT generate or write any test files or expected values JSON.\n"
        # ... additional guardrails and JSON output schema omitted for brevity ...
    ),
)
\end{lstlisting}

\subsubsection*{Translation Agent}

The \texttt{translation\_agent}, also powered by GPT-5.4, performs the core R-to-Python translation. Its system prompt specialises it for actuarial computations and encodes explicit mappings between common R and Python idioms -- for example, \texttt{data.frame} to \texttt{pandas DataFrame}, \texttt{glm()} to \texttt{statsmodels GLM}, and 1-based to 0-based indexing -- so that domain conventions are preserved across the translation.

Crucially, the translation agent participates in the system's feedback loop. When compilation or test failures are routed back, it receives the error tracebacks, reads the current translated file, and applies targeted fixes rather than rewriting the entire file from scratch -- a strategy inspired by the self-debugging paradigm~\cite{chen2024selfdebugging}. The following snippet shows the initialisation with an abbreviated system prompt.

\begin{lstlisting}[language=]
# -- Agent 2: Translation Agent --
translation_agent = create_agent(
    model="openai:gpt-5.4-2026-03-05",
    tools=[read_file, split_r_code, write_translated_file],
    name="translation_agent",
    system_prompt=(
        "You are an expert in both R and Python, specializing in actuarial computations.\n"
        "Your job is to translate R code into equivalent Python code.\n\n"
        "TRANSLATION GUIDELINES:\n"
        "- R data.frame -> pandas DataFrame\n"
        "- R matrix operations -> numpy arrays\n"
        "- R dplyr (filter, mutate, group_by, summarise) -> pandas equivalents\n"
        "- R library(DBI)/RSQLite -> sqlite3 or sqlalchemy\n"
        "- R glm() -> statsmodels GLM\n"
        "- R's 1-based indexing -> Python's 0-based indexing\n"
        # ... additional mapping rules and workflow details omitted for brevity ...
        "If you receive feedback about compilation errors or test failures:\n"
        "1. FIRST read the error traceback carefully. Identify the exact line number,\n"
        "   exception type, and error message.\n"
        "2. Use read_file to read the CURRENT translated file to see what's at that line.\n"
        "3. Fix the specific issue -- do NOT rewrite the entire file from scratch unless\n"
        "   the error is fundamental. Targeted fixes are more reliable.\n"
        "4. Call write_translated_file exactly once with the corrected code, then\n"
        "   produce a brief final response and stop.\n"
        # ... further guardrails against touching test artifacts omitted ...
    ),
)
\end{lstlisting}

\subsubsection*{Compilation, Test Runner, and Report Agents}

The \texttt{compilation\_agent} and \texttt{test\_runner\_agent} are both powered by GPT-5.4 mini, reflecting the comparatively simpler nature of their tasks. The compilation agent validates Python syntax and executes the translated file, reporting a structured pass/fail status with full error tracebacks on failure. The test runner agent executes the pre-written test suite and captures per-test metadata (name, category, pass/fail status), producing a structured JSON report. Both agents serve as automated quality gates, providing the translation agent with actionable feedback when failures occur.

The \texttt{report\_agent} (GPT-5.4 mini) generates a comprehensive Markdown migration report following a fixed six-section template covering the original R code's functionality, the translation approach, agent execution statistics, challenges encountered, test results, and files produced. It reads the agent execution log to include pre-computed summary statistics rather than generating them from scratch. The complete initialisation code for these three agents is available in the Jupyter notebook.

\subsection{Results}

We evaluated the legacy code migration MAS on two actuarial reserving examples of increasing complexity, both implemented in R and accompanied by synthetic data. Complete source code, translations, and data files are available in the GitHub repository.

\begin{itemize}
  \item \textbf{Simple: Chain-Ladder Reserving} (198 lines of R code): Reads a $10 \times 10$ incremental claims triangle from CSV, converts to cumulative form, calculates volume-weighted age-to-age development factors~\cite{mack1993distribution}, derives cumulative development factors to ultimate, and projects unpaid claims reserves by origin year. The pre-written test suite comprises 14 tests (6~data/format, 8~content/numerical).

  \item \textbf{Difficult: GLM-Based Reserving with Bootstrap} (265 lines of R code): Loads a $15 \times 15$ claims triangle from CSV and policy/premium data from a SQLite database, fits an over-dispersed Poisson GLM~\cite{england2002stochastic}, predicts the lower triangle, and performs 1,000 bootstrap iterations with process variance to estimate reserve distributions. The pre-written test suite comprises 15 tests (8~data/format, 7~content/numerical).
\end{itemize}

Table~\ref{tab:chain_ladder_tests} summarises the 14 pre-written, R-verified tests that the translated Python code must satisfy for the simple chain-ladder example. Reference values were computed by executing the original R script with full numerical precision and are stored in a separate JSON file that is deliberately isolated from the translation agent's context.

\begin{table}[h!]
\centering
\footnotesize
\caption{Pre-written test suite for the chain-ladder reserving translation. Each test is classified as either a \emph{data/format} test (verifying structural properties) or a \emph{content/numerical} test (verifying computed values against R ground truth).}
\label{tab:chain_ladder_tests}
\begin{tabular}{@{}clp{8.5cm}@{}}
\toprule
\textbf{\#} & \textbf{Category} & \textbf{Description} \\
\midrule
1  & Data/Format       & Triangle CSV loads successfully with shape $(10, 10)$ \\
2  & Data/Format       & Column names are \texttt{dev\_1}\,...\,\texttt{dev\_10}; index labels are 2010--2019 \\
3  & Data/Format       & All development columns have numeric dtype \\
4  & Data/Format       & Missing values follow upper-triangle pattern (contiguous observations per row) \\
5  & Data/Format       & At least one public entry point (\texttt{run\_chain\_ladder}, \texttt{chain\_ladder}, or \texttt{main}) is callable \\
6  & Data/Format       & Model returns a supported container type (dict, DataFrame, or tuple) \\
\midrule
7  & Content/Numerical & Cumulative values for origin year 2010 match R output (10 values) \\
8  & Content/Numerical & Volume-weighted development factors match R output (9 values) \\
9  & Content/Numerical & Number of valid pairs per development step matches R (9 counts) \\
10 & Content/Numerical & Cumulative development factors to ultimate match R output (10 values) \\
11 & Content/Numerical & Latest observed cumulative values and development indices match R (10+10 values) \\
12 & Content/Numerical & Projected ultimate claims and unpaid reserves per origin year match R output (10+10 values) \\
13 & Content/Numerical & Total unpaid claims reserve matches R output ($15{,}316.34$) \\
14 & Content/Numerical & Result table contains expected origin-year and latest-period labels \\
\bottomrule
\end{tabular}
\end{table}

To assess robustness and correctness, we executed each example 10 times end-to-end. Table~\ref{tab:migration_results} summarises the results. Both examples achieved a 100\% pass rate across all 10 runs. The chain-ladder example completed with negligible retry effort (mean 0.2, median 0, range 0--1 retries), reflecting its deterministic nature and modest complexity. The GLM-based example required more iterative refinement (mean 1.2, median 1, range 0--4 retries); in particular, one of the 10 GLM runs needed 4 retries before all tests passed, indicating that the feedback loop can absorb several rounds of error correction before the pipeline succeeds.

\begin{table}[h!]
\centering
\footnotesize
\caption{Summary of migration results across 10 repeated end-to-end runs per example.}
\label{tab:migration_results}
\setlength{\tabcolsep}{5pt}
\begin{tabular}{@{}lcccc@{}}
\toprule
\textbf{Example} & \textbf{R Lines} & \textbf{\shortstack{Tests\\(Data / Content)}} & \textbf{Pass Rate} & \textbf{\shortstack{Mean / Median Retries\\(Range)}} \\
\midrule
Chain-Ladder Reserving    & 198 & 14 (6 / 8) & 100\% & 0.2 / 0 (0--1) \\
GLM-Based Reserving       & 265 & 15 (8 / 7) & 100\% & 1.2 / 1 (0--4) \\
\bottomrule
\end{tabular}
\end{table}

For the simple chain-ladder example, whose outputs are fully deterministic -- development factors, cumulative development factors, and reserve totals -- the numerical results were identical across all 10 runs to floating-point precision. The more complex GLM-based example combines deterministic outputs (including the GLM dispersion parameter and reserve-to-premium ratios) with stochastic outputs (bootstrap distributions): the deterministic values were again identical across all 10 runs, while for the stochastic outputs the shape, type, and sanity tests passed in every run and the numerical values varied naturally across runs due to random sampling, as expected. In both examples, the Python code translations themselves exhibited slight variations in variable naming, code structure, and inline comments between runs -- a natural consequence of LLM non-determinism -- but all 10 runs produced functionally equivalent code that passed the complete test suite.

The ground-truth testing strategy relies on reference values pre-computed from the original R code with full numerical precision, stored in a separate JSON file, and deliberately isolated from the translation agent's context. This ensures that validation is independent of the translation process and identical across runs. Each test suite combines structural assertions (verifying data shape, schema, and callability) with numerical assertions checked against R-derived reference values, providing a structured factuality audit of the translated code.

\subsection{Implications for Actuarial Practice}
\label{sec:mas_risks}

This case study demonstrates that a multi-agent system can automate end-to-end migration of actuarial R code to Python with ground-truth validation, decomposing a complex task into specialised stages linked by a sequential workflow with explicit feedback loops. The modular, test-gated design provides a level of assurance that goes beyond manual code review and extends naturally to other source--target language pairs -- such as R, SAS, COBOL, or VBA migrated to modern targets like Python, Java, or C\# -- by adapting agent prompts and tools. More broadly, the MAS paradigm itself is applicable wherever actuarial workflows decompose into discrete, coordinated subtasks, with different design patterns (hardcoded sequential, supervisor-based, hierarchical, or decentralised peer architectures) entailing different trade-offs between control and flexibility~\cite{pan2024multiagent}.

Code translation, however, is typically only one stage of a broader migration pipeline. A comprehensive migration effort may additionally involve creating a catalogue of existing models, data, and their dependencies, enriching legacy code with contextual metadata, performing dependency and impact analysis to determine migration sequencing, and establishing parallel-run validation environments where legacy and migrated systems are compared over extended periods. The context enrichment stage in particular requires deep actuarial expertise that cannot be fully automated. The modular architecture presented here is designed to be composable: individual agents or entire sub-workflows can be embedded as components within a larger orchestration framework that coordinates the end-to-end process. The growing availability of specialised coding agents -- such as Claude Code\footnote{\url{https://docs.anthropic.com/en/docs/claude-code}}, OpenAI's Codex\footnote{\url{https://openai.com/index/introducing-codex/}}, Cursor\footnote{\url{https://www.cursor.com/}}, and Windsurf\footnote{\url{https://windsurf.com/}} -- and interoperability standards such as the Model Context Protocol (MCP)\footnote{\url{https://www.anthropic.com/news/model-context-protocol}} further lower the adoption barrier, since such tools can serve either as standalone assistants or as components within larger architectures.

Despite these advantages, deploying LLM-based MAS in regulated insurance contexts requires careful risk management. LLM-based translations vary between runs, complicating audit trails; LLMs may produce silently incorrect outputs that pass tests but introduce subtle errors in untested code paths~\cite{yetistiren2023evaluating}; proprietary actuarial code raises confidentiality concerns when processed by external APIs; and source material (including comments) can in principle be crafted to manipulate agent behaviour through prompt injection~\cite{owasp2025llmtop10}. These and related concerns -- including governance controls for non-determinism, drift, human oversight, and escalation -- are discussed in Section~\ref{sec:challenges_and_considerations}. For production use, human review of translated code, comprehensive test suites with both structural and numerical assertions, and escalation procedures for runs that exhaust the maximum retry count remain essential.

As with the other case studies in this article, human--AI collaboration is central: translated code should be reviewed by domain experts before deployment, test suites should be inspected for coverage, and migration reports should be validated against business intent. In more advanced deployments, such oversight can be embedded programmatically through human-in-the-loop approval nodes at critical decision points. Taken together, the results suggest that MAS-based migration can complement -- rather than replace -- deterministic tools and human judgement.

%% file: further_applications.tex
\section{Further Applications of Generative AI in Actuarial Science}
\label{sec:further_applications}

The four case studies presented in the preceding sections illustrate specific, implemented applications of generative AI in actuarial workflows. However, the potential scope of GenAI extends well beyond these examples. Drawing on an insurance value chain framework~\cite{eling2022impact} and a recent systematic review of AI in insurance~\cite{bhattacharya2025ai}, this section surveys additional application areas, organised by their proximity to policyholders and the associated level of risk and regulatory scrutiny. For each area, we highlight representative opportunities and, where relevant, the practical barriers that currently limit adoption.

\subsection{Consumer-Facing Applications}

GenAI offers opportunities to improve direct interactions with policyholders, though these applications carry elevated reputational and regulatory risk. LLM-driven chatbots with advanced text, voice, and vision capabilities can address customer queries, retrieve policy details, and clarify coverage information with a degree of personalisation that is difficult to achieve with rule-based systems. In claims processing, GenAI can assist with digitising, classifying, and routing incoming documents, and can generate draft settlement recommendations to accelerate workflows. Policy renewal optimisation -- analysing renewal trends and generating tailored recommendations -- represents a further consumer-adjacent application where the balance between personalisation and the risk of perceived intrusiveness must be carefully managed.

However, insurers must weigh these opportunities against the risk of generating incorrect, misleading, or inappropriate responses that reach policyholders directly. Errors in consumer-facing applications can erode trust, attract regulatory attention, and cause reputational damage that is difficult to reverse. In practice, many insurers currently restrict GenAI deployment to internal, non-consumer-facing applications, and pilot consumer-facing use cases only with human review in the loop~\cite{bhattacharya2025ai}.

\subsection{Internal and Analytical Applications}

Applications that support internal actuarial and operational workflows present a lower risk profile and are generally more amenable to near-term adoption. Automated reporting is a prominent example: GenAI can draft regulatory reports such as the Solvency and Financial Condition Report (SFCR) or internal risk reports by leveraging historical document versions -- either through few-shot prompting, where prior reports are supplied as in-context examples, or through fine-tuning -- combined with retrieval from databases and document repositories, and tool use for numerical calculations, though human review of all generated content remains essential for compliance.

In underwriting, GenAI can analyse applicant data from multiple sources and generate structured summaries for risk evaluation, reducing the time required for manual review. Product development and pricing workflows can benefit from LLMs that draft policy terms, perform comparative analyses, and -- through function calling and code interpreters -- execute pricing calculations. In strategic scenario modelling, GenAI can help actuaries brainstorm and articulate narrative stress-test scenarios for capital management and Own Risk and Solvency Assessment (ORSA) -- for instance, compound events combining a pandemic with a cyberattack or an interest-rate shock with a catastrophic weather event -- which subject-matter experts then translate into quantitative assumptions and validate for plausibility and internal consistency. Finally, employee training and knowledge dissemination can be supported through interactive LLM-based systems that adapt explanations to individual knowledge levels, complementing the broader productivity gains from GenAI-assisted professional writing observed in controlled experiments~\cite{noy2023experimental}.

Taken together, these examples illustrate the breadth of opportunities that GenAI presents for actuarial practice, while underscoring that each application carries its own risk profile -- including a dual-use dimension, whereby the same capabilities that support legitimate actuarial tasks can be repurposed to facilitate fraud or manipulation -- and requires careful evaluation before deployment. The following section discusses these cross-cutting challenges and governance considerations in detail.

%% file: challenges_and_considerations.tex
\section{Risks and Governance of GenAI in Insurance}
\label{sec:challenges_and_considerations}

The case studies and applications discussed in the preceding sections demonstrate the potential of Generative AI to support actuarial work. However, deploying GenAI in regulated insurance environments introduces risks that must be identified, assessed, and mitigated before any production use. This section provides a structured, risk-centred discussion of the key challenges, organised around the threat categories most relevant to actuarial and insurance deployment. A summary of risks, mitigations, and supporting references is provided in Table~\ref{tab:risk_summary}.

\subsection{Regulatory and Legal Compliance}

The regulatory landscape for AI in insurance is evolving rapidly. The European Union (EU) Artificial Intelligence Act~\cite{euaiact2024} establishes a risk-based framework that classifies AI systems by their potential for harm. Insurance-related applications may fall into the high-risk category under the Act, particularly where AI systems are used for risk assessment and pricing decisions in life and health insurance that affect individuals' access to coverage or the terms on which it is offered. The Digital Operational Resilience Act (DORA) imposes additional requirements on information and communications technology (ICT) risk management for financial institutions, including insurers. The governance principles proposed by the European Insurance and Occupational Pensions Authority (EIOPA) provide a sector-specific foundation for ethical and trustworthy AI deployment~\cite{EIOPA2021AIprinciples}. Complementary guidance is provided by the U.S. National Institute of Standards and Technology (NIST) AI Risk Management Framework~\cite{nist2023airmf} and its GenAI-specific profile~\cite{nist2024genai}, alongside a detailed analysis of the EU AI Act's implications for credit underwriting and insurance~\cite{hacker2025euaiact}.

A core regulatory challenge is the non-deterministic nature of LLM outputs, which complicates the auditability and reproducibility that regulators expect (see Section~\ref{sec:nondeterminism} for a detailed discussion). Addressing this requires concrete production controls rather than general acknowledgement. Key controls that merit consideration include: (i)~acceptance criteria that define when an LLM output is considered valid for a given use case, (ii)~human-in-the-loop review thresholds that specify which outputs require manual approval, (iii)~comprehensive logging of all prompts, model versions, context windows, and outputs to support retrospective audit, and (iv)~rollback and fallback procedures that allow reverting to non-AI workflows when model behaviour deviates from expectations. These controls should be embedded in an organisation's model risk management framework and reviewed periodically as regulatory guidance matures.

\subsection{Security Threats and Adversarial Risks}

GenAI systems introduce novel attack surfaces that differ from those of traditional software. Prompt injection -- in which adversarial input manipulates an LLM into behaviours such as ignoring its instructions or producing unintended outputs -- represents a well-documented and partially unresolved threat~\cite{owasp2025llmtop10, liu2024promptinjection}. In actuarial applications, prompt injection could occur through user inputs, through documents processed by RAG pipelines (a variant known as retrieval poisoning), or -- as discussed in Section~\ref{sec:mas_risks} -- through legacy source code fed to coding agents. Jailbreaking attacks, in which carefully crafted prompts bypass safety filters, pose additional risks when LLMs are exposed to end users or external data.

Mitigations include input validation and sanitisation of all content processed by LLMs, output filtering to detect anomalous or policy-violating responses, execution sandboxing for code-generating agents, and rate limiting to constrain the scope of potential exploitation. Defence-in-depth strategies that combine multiple layers of protection are preferable to relying on any single safeguard~\cite{bommasani2022foundation}.

\subsection{Dual-Use and Fraud Enablement}

Many GenAI capabilities have a dual-use dimension. The same vision-enabled models that detect anomalous patterns in claims images can generate or modify images to fabricate damage evidence. LLMs that draft policy documents or assess claims narratives can equally produce convincing fraudulent documentation. The accessibility of GenAI tools lowers the barrier to entry for sophisticated fraud schemes that previously required specialised technical skills~\cite{mitra2024, swissre2025sonar}.

Insurers deploying GenAI should therefore invest in countermeasures alongside adoption. These include digital provenance and watermarking for AI-generated content, forensic analysis tools that detect synthetic media, strict capture protocols for evidence submission (e.g., requiring metadata-rich, time-stamped photographs), and clear escalation rules that route suspicious content to specialised investigation teams. The development of detection capabilities should proceed in parallel with the deployment of generative capabilities to avoid creating an asymmetry that favours attackers.

\subsection{Non-Determinism, Drift, and Reproducibility}
\label{sec:nondeterminism}

LLMs are inherently non-deterministic: even with identical inputs, outputs may vary due to sampling parameters, floating-point arithmetic, and infrastructure-level differences. This poses challenges for actuarial applications where reproducibility is a professional and regulatory expectation. As demonstrated in Case Study~4 (Section~\ref{sec:multi_agent_system}), repeated executions of the same migration task produced functionally equivalent but textually different code translations across 10 runs per example. While all translations passed the automated test suite, the variation complicates traditional notions of audit trails and version control.

Model drift -- changes in behaviour resulting from provider-side model updates -- represents a related concern. An LLM-based pipeline that performs reliably with one model version may produce degraded or different outputs after a silent backend update. Concrete mitigations include: version-pinning LLM endpoints to specific model version identifiers, maintaining regression test suites that are re-executed after any model change, controlling sampling parameters (temperature, seed) where supported by the API, and documenting the model version and configuration alongside every production output. For critical actuarial applications, organisations should evaluate whether the inherent non-determinism of LLMs is acceptable or whether deterministic alternatives (e.g., rule-based systems, traditional NLP) are more appropriate for the specific task.

\subsection{Privacy, Data Leakage, and Confidentiality}

Actuarial workflows frequently involve personal data (e.g., policyholder information, health records, claims histories), proprietary business logic (e.g., pricing models, reserving methodologies), and confidential corporate information (e.g., financial projections, strategic plans). Sending such data to external LLM APIs operated by third parties creates exposure to data leakage, unauthorised retention, and potential use of submitted data for model training -- even where providers contractually disclaim such use, enforcement and verification remain challenging. These providers also wield considerable power: a handful of dominant vendors unilaterally set model architectures, data handling, and commercial terms, leaving insurers with little leverage.

The General Data Protection Regulation (GDPR) imposes strict requirements on the processing of personal data, including purpose limitation, data minimisation, and the right to erasure, all of which are difficult to guarantee when data passes through opaque LLM infrastructure. Mitigations include deploying on-premise or private-cloud LLM instances that keep data within the organisation's control boundary, anonymising or pseudonymising personal data before LLM processing, establishing data processing agreements with API providers that specify retention limits and usage restrictions, and conducting data protection impact assessments for each GenAI use case.

\subsection{Governance, Monitoring, and Human Oversight}

Effective governance of GenAI in insurance requires structures that go beyond traditional IT governance. Drawing on the EIOPA principles for AI governance~\cite{EIOPA2021AIprinciples}, the International Actuarial Association's AI governance framework~\cite{IAA2025governance}, the American Academy of Actuaries' professionalism guidance on generative AI~\cite{aaa2024genaiprof}, and the broader literature on AI risk in insurance~\cite{eling2022impact}, several governance controls can be identified as particularly relevant for regulated deployment. First, human oversight at decision points: GenAI outputs that influence policyholder-affecting decisions (e.g., underwriting, claims, pricing) should be reviewed by qualified personnel before action is taken, with the level of oversight proportionate to the risk and reversibility of the decision. Second, monitoring and alerting: production GenAI systems should be instrumented to track output quality metrics, latency, error rates, and content policy violations, with automated alerts triggered when metrics deviate from established baselines. Third, escalation procedures: clear escalation paths should exist for cases where GenAI outputs are ambiguous, potentially harmful, or outside the system's validated operating envelope, directing such cases to human experts with domain authority. Fourth, audit logging: all interactions with GenAI systems -- including prompts, retrieved context, model responses, and human review decisions -- should be logged immutably to support retrospective audit and regulatory examination. Fifth, periodic review and revalidation: GenAI systems should be subject to periodic revalidation against their original acceptance criteria, particularly after model updates, changes in the operating environment, or regulatory guidance updates.

\subsection{Organisational and Practical Barriers}

Beyond technical and regulatory risks, several organisational factors influence the feasibility and pace of GenAI adoption in insurance. Reputational risk is a primary concern: an AI-generated error that reaches a policyholder or regulator can attract disproportionate public attention relative to an equivalent human error, reflecting heightened societal scrutiny of automated decision-making. Consumer trust in AI-driven processes remains low in many markets, and insurers must invest in transparency and communication to build acceptance.

The expertise required to design, implement, and maintain GenAI systems -- spanning prompt engineering, LLM evaluation, domain-specific validation, and ongoing monitoring -- represents a barrier for many actuarial teams. Interdisciplinary collaboration among actuaries, data scientists, AI engineers, and compliance specialists is highly beneficial but difficult to organise in practice. The financial costs of LLM API usage, computational infrastructure, and the energy consumption of large-scale model inference add further burdens, particularly for smaller insurers. Investing in education and skill-building programmes for actuaries and other insurance professionals is critical to closing the knowledge gap and enabling informed participation in GenAI governance decisions.

\begin{table}[h!]
\centering
\footnotesize
\caption{Summary of key risks associated with GenAI deployment in actuarial and insurance contexts, together with recommended mitigations and supporting references.}
\label{tab:risk_summary}
\begin{tabular}{p{2.3cm} p{3.4cm} p{3.8cm} p{1.6cm}}
\toprule
\textbf{Risk Category} & \textbf{Key Threats} & \textbf{Recommended Mitigations} & \textbf{References} \\
\midrule
\raggedright Regulatory and Legal Compliance & \raggedright Non-deterministic outputs; evolving AI regulation & \raggedright Acceptance criteria; logging; rollback procedures & \cite{euaiact2024, EIOPA2021AIprinciples, nist2023airmf, hacker2025euaiact} \\
\addlinespace
\raggedright Security Threats and Adversarial Risks & \raggedright Prompt injection; retrieval poisoning; jailbreaking & \raggedright Input validation; output filtering; sandboxing & \cite{owasp2025llmtop10, bommasani2022foundation, liu2024promptinjection} \\
\addlinespace
\raggedright Dual-Use and Fraud Enablement & \raggedright Synthetic media; fabricated documents; convincing narratives & \raggedright Provenance checks; forensic tools; capture protocols & \cite{mitra2024, swissre2025sonar} \\
\addlinespace
\raggedright Non-Determinism, Drift, and Reproducibility & \raggedright Output variability; silent model updates & \raggedright Version pinning; regression testing; seed control & \cite{nist2024genai} \\
\addlinespace
\raggedright Privacy, Data Leakage, and Confidentiality & \raggedright Data sent to external APIs; GDPR exposure & \raggedright On-premise deployment; anonymisation; data protection impact assessments & \cite{EIOPA2021AIprinciples} \\
\addlinespace
\raggedright Governance, Monitoring, and Human Oversight & \raggedright Insufficient oversight; missing audit trails & \raggedright Human-in-the-loop; monitoring; immutable logging & \cite{eling2022impact, EIOPA2021AIprinciples, IAA2025governance, aaa2024genaiprof} \\
\addlinespace
\raggedright Organisational and Practical Barriers & \raggedright Reputational risk; skill gaps; cost; consumer trust & \raggedright Training programmes; interdisciplinary teams; transparency & \cite{eling2022impact, noy2023experimental} \\
\bottomrule
\end{tabular}
\end{table}

\subsection{Minimum Controls by Case Study}

The controls described above apply across GenAI deployments, but their relative priority depends on the data and workflow of each application. Table~\ref{tab:case_study_controls} distils, for the four case studies of this article (CS1--CS4; Sections~\ref{sec:case_study_crash_reports}--\ref{sec:multi_agent_system}), the \emph{minimum} controls we consider necessary before any production use. These are not exhaustive -- the broader set of recommended mitigations is summarised in Table~\ref{tab:risk_summary} -- but they capture the baseline safeguards most directly implied by each case study's dominant risk.

\begin{table}[h!]
\centering
\footnotesize
\caption{Minimum production controls mapped to each case study, instantiated from the risk categories of this section (Table~\ref{tab:risk_summary}) for the specific data and workflow of each case study.}
\label{tab:case_study_controls}
\begin{tabular}{@{}p{2.2cm} p{3.3cm} p{5.8cm}@{}}
\toprule
\textbf{Case Study} & \textbf{Dominant risk exposure} & \textbf{Minimum controls before production} \\
\midrule
CS1: Claim-cost prediction from claims text & Personal/health data in free-text claims; extraction quality; to a lesser degree, discrimination (proxy bias via derived features) and dual-use fraud (AI-fabricated claim narratives) & Pseudonymise or redact personal data before extraction (or use a private/on-premise model); define acceptance criteria and human spot-checks for the extracted fields; assess the derived features and predictions for proxy discrimination and screen claim narratives for fraud indicators; pin the model version and cache raw completions \\
\addlinespace
CS2: Market comparison via RAG & Retrieval/generation errors (hallucination); public-report data, hence limited privacy exposure & Validate every extracted figure against the source report; expose retrieved chunks for human review and traceability; pin model versions and retrieval parameters \\
\addlinespace
CS3: Car-damage classification (fine-tuned vision) & Dual-use/fraud via synthetic or tampered images; personal data in images & Enforce capture protocols (metadata-rich, time-stamped images) with synthetic-media and provenance checks; strip metadata and redact identifying content; require human review of classifications; revalidate after model updates \\
\addlinespace
CS4: Legacy code migration (multi-agent system) & Untrusted code execution; prompt injection via legacy code; confidentiality of proprietary code & Execute agent-generated code only in a sandbox; gate each migration on a regression test suite checking functional equivalence; require human code review and escalation when retries are exhausted; keep proprietary code within the organisation's boundary \\
\bottomrule
\end{tabular}
\end{table}

%% file: conclusion.tex
\section{Conclusion}
\label{sec:conclusion}

This article examined the potential of generative AI to support actuarial science and the broader insurance industry through four implemented case studies. In Case Study~1, LLMs extracted structured variables from free-form claim descriptions, improving the predictive performance of a downstream machine learning model. Case Study~2 demonstrated that Retrieval-Augmented Generation can extract, align, and synthesise financial and regulatory data from insurers' annual reports, converting a labour-intensive manual process into an automated pipeline. Case Study~3 showed that vision-enabled LLMs can process vehicle images directly for damage classification and localisation, with the off-the-shelf model already reaching classification performance comparable to a conventional CNN baseline and fine-tuning yielding a further improvement over the non-fine-tuned version. Case Study~4 introduced a multi-agent system that can autonomously migrate actuarial legacy code from R to Python, validating each translation against a pre-written test suite that combines structural checks with content-level numerical checks against ground-truth values computed from the original R code.

Across these case studies, generative AI improved predictive modelling through LLM-derived features, automated structured data extraction from unstructured documents, enabled image-based damage assessment, and supported code migration with automated validation. At the same time, our analysis highlights that responsible deployment requires careful attention to the operational and governance risks that accompany these capabilities: non-determinism and reproducibility challenges, security threats including prompt injection, dual-use and fraud enablement concerns, privacy and data leakage, and the need for robust governance controls tailored to regulated insurance environments. The gap between a working prototype and a production-ready system is substantial, and organisations should invest in monitoring, human oversight, and periodic revalidation alongside technical development.

As AI becomes increasingly embedded in actuarial workflows, the profession must balance the adoption of tools that demonstrably improve practice with the standards of transparency, reproducibility, and accountability that regulators and policyholders expect. With this article, we aim to equip actuaries with a clearer understanding of where generative AI delivers value, where its limitations and risks must be managed, and how to critically assess its use in their own work.